\documentclass{JHEP3}

\usepackage{Pformel,Pmeta}
\usepackage{amsmath,amsfonts,amsthm,times}

\newcommand{\ev}{U}

\preprint{PTA/08-068}

\title{On $s\ell_3$ Knizhnik--Zamolodchikov equations \\ and ${\cal W}_3$ null-vector equations}

\author{Sylvain Ribault
\\ 
 Laboratoire de Physique Th\'eorique et Astroparticules, UMR5207 CNRS-UM2,
 \\
 Universit\'e Montpellier II, Place E. Bataillon, CC 070
 \\
 34095 Montpellier Cedex 05, France 
 \\
 {\footnotesize \tt ribault@lpta.univ-montp2.fr }
}

\abstract{Starting from Sklyanin's separation of variables for the $s\ell_3$ Yangian model, we derive the separation of variables for the quantum $s\ell_3$ Gaudin model. We use the resulting new variables for rewriting the 
$s\ell_3$ Knizhnik--Zamolodchikov equations, and comparing them with certain null-vector equations in conformal field theories with ${\cal W}_3$-algebra symmetry. The two sets of equations are remarkably similar, but become identical only in the critical level limit. This is in contrast to the $s\ell_2$ Knizhnik--Zamolodchikov equations, which are known to be equivalent to Belavin--Polyakov--Zamolodchikov equations for all values of the level. }

\makeatletter\let\default@color\current@color\makeatother 

\begin{document}

\zeq\section{Introduction and conjecture}

Many interesting models of two-dimensional conformal field theories are based on affine Lie algebras $\widehat{s\ell_N}$ and their cosets, starting with Wess--Zumino--Witten models. To solve such theories is an interesting challenge, whose difficulty depends more from the choice of the underlying Lie algebra $s\ell_N$, than from the particular coset or real form chosen. 

For example, the $s\ell_2$ family includes string theory in $AdS_3$ and in the $\SLU$ 2d black hole, as well as the $\Hp$ model; the simplest non-rational nontrivial model of the family is however Liouville theory, also known as conformal $s\ell_2$ Toda theory. In several of the other theories in the $s\ell_2$ family, it turns out that arbitrary correlation functions have a simple relation to certain Liouville theory correlation functions \cite{rt05,hs07}. This relation entails a relation between the Knizhnik--Zamolodchikov equations which follow from $\widehat{s\ell_2}$ symmetry, and the Belavin--Polyakov--Zamolodchikov equations which follow from the conformal symmetry of Liouville theory \cite{sto00}. The relation to Liouville theory is helpful in solving certain models in the $s\ell_2$ family, by disentangling the particular details of a model from its general $s\ell_2$-based properties. For example, the $\Hp$-Liouville relation was very helpful in solving the $\Hp$ model on a disc \cite{hr06}. Moreover, playing with the Liouville side of the relation leads to the discovery of new conformal field theories which generalize the $\Hp$ model \cite{rib08}, and which can be considered as members of an extended $s\ell_2$ family. 

The intuitive reason why such a relation exists is that $\widehat{s\ell_2}$ representations are parametrized by just one number, their spin. So it is not very surprising that the dynamics of say the $\Hp$ model, a theory of three interacting bosons, are in some sense effectively one-dimensional. Applied to a theory with an $\widehat{s\ell_{N>2}}$ symmetry algebra, which may involve as many as $N^2-1$ bosons, this reasoning suggests that it could be related to a theory of only $N-1$ bosons. Such a theory is present in the $s\ell_N$ family: namely, conformal $s\ell_N$ Toda theory, which can be described by the Lagrangian $L= (\p \phi,\bp \phi) +\sum_{i=1}^{N-1} e^{b(e_i,\phi)}$ where the field $\phi(z,\bz)$ and the simple roots $e_i$ live in the $N-1$-dimensional root space of $s\ell_N$. (See for example \cite{fl07c} for details.) 
It is therefore natural to investigate whether correlation functions of that theory have a simple relation to correlation functions of other models in the family. Such a relation would be a welcome simplification: for instance, in the $s\ell_3$ family, we would trade 8 bosons of the $SL(3,\R)$ WZW model for the 2 bosons of $s\ell_3$ conformal Toda theory. 

The investigation of the $s\ell_{N>2}$ families is motivated both from the appearance of groups of rank higher than one in many interesting string theory backgrounds, and from the observation that theories in the $s\ell_{N>2}$ families are qualitatively more difficult, and more generic, than theories in the $s\ell_2$ family. This is due to features like: infinite fusion multiplicities, correlation functions involving degenerate fields without obeying nontrivial differential equations, and structures constants which can probably not be written in terms of known special functions \cite{fl07c}. These are serious obstacles in the way of solving such theories. Nevertheless, we do know a strong explicit constraint on the correlation functions of all models which have the full $\widehat{s\ell_N}$ symmetry: they obey KZ equations. The aim of the present article is therefore to determine whether the $s\ell_3$ KZ equations are related to some null-vector equations in conformal $s\ell_3$ Toda theory, which follow from its symmetry algebra ${\cal W}_3$. 

In analogy with the $s\ell_2$ case, we will look for a relation based on Sklyanin's separation of variables \cite{skl92c}. As the KZ equations are closely related to the Gaudin Hamiltonians, we will use Sklyanin's separation of variables for the quantum $s\ell_3$ Gaudin model. Before using it, we will actually have to work it out, as this has apparently not been fully done in the existing literature. A rather close starting point is available though: the separation of variables for the $s\ell_3$ Yangian model \cite{skl92b}.

Let us now sketch the correlation functions we are interested in and the relation we are aiming at. 
Consider a theory with an $\widehat{s\ell_3}$ symmetry algebra at level $k$. 
We are interested in correlation functions of generic $\widehat{s\ell_3}$ affine primary fields $\Phi^j(x|z)$, where the spin $j$ labels $s\ell_3$ representations, the variable $x$ is a generic isospin coordinate (a triplet of complex numbers), and $z$ is a coordinate on the complex plane where the field lives. We denote an $n$-point function of such fields as  
\bea
\Omega_n \equiv \la \prod_{i=1}^n \Phi^{j_i}(x_i|z_i)\ra\ .
\label{omn}
\eea
We will seek to relate such correlation functions to fairly particular correlation functions in a theory with a ${\cal W}_3$ symmetry algebra at parameter $b=(k-3)^{-\frac12}$, which involve not only $n$ generic ${\cal W}_3$-primary fields $V_{\al_i}(z_i)$ corresponding to $\Phi^{j_i}(x_i|z_i)$, but also $3n-6$ degenerate fields $V_{-b^{-1}\om_1}(y_a)$ with the special value $-b^{-1}\om_1$ for their ${\cal W}_3$ momentum:
\bea
\tilde{\Omega}_n \equiv \la  \prod_{a=1}^{3n-6} V_{-b^{-1}\om_1}(y_a)\prod_{i=1}^n V_{\al_i}(z_i) \ra \ .
\label{tomn}
\eea
The number of degenerate fields is of the order of $3n$, which allows their worldsheet positions $y_a$ to (approximately) correspond to the $3n$ components of the isospin variables $x_1\cdots x_n$. This will also allow
$\tilde{\Omega}_n$ to obey some differential equations which may be related to the KZ equations for $\Omega_n$. 
Moreover, the tentative relation between $\Omega_n$ and $\tilde{\Omega}_n$ will involve a simple twist function
\bea
\Theta_n = \prod_{a< b} (y_a-y_b)^\lambda \prod_i\prod_a (y_a-z_i)^\mu \prod_{i< j}(z_i-z_j)^\nu\ ,
\label{thn}
\eea
for some constants $\lambda,\mu,\nu$ to be determined in terms of the level $k$ of our $\widehat{s\ell_3}$ algebra; and the integral transformation ${\cal K}$ with integration kernel $K(\{x_i\}|\{y_a\},\ev|\{z_i\})$ which implements Sklyanin's separation of variables, and may therefore depend on the spins $j_i$ but not on the level $k$. We will then investigate the validity of the conjecture $\Omega_n\overset{?}{\sim} {\cal K}\cdot \Theta_n \tilde{\Omega}_n \equiv \int d\ev \prod_a dy_a\ \ K\cdot \Theta_n \tilde{\Omega}_n$, or more explicitly
\bea
\boxed{
\Omega_n(\{x_i\}|\{z_i\}) \overset{?}{\sim}  \int \! d\ev\prod_a dy_a\ \ K(\{x_i\}|\{y_a\},\ev|\{z_i\})\cdot \Theta_n(\{y_a\}|\{z_i\})\tilde{\Omega}_n(\{y_a\}|\{z_i\})\ .
}
\label{omom}
\eea
The meaning of the equivalence $\sim$ here is that both sides obey the same differential equations. If true, this equivalence may then be promoted to a relation between physical correlation function of specific models, like the relation between the $\Hp$ model and Liouville theory \cite{rt05}, but this is not the focus of the present article. This is why we do not worry about such details as the dependence of the correlation functions on antiholomorphic variables. 

The article will start with a brief review of the KZ equations and other Ward identities in conformal field theories with $\widehat{s\ell_N}$ symmetries, where we will explain how the Gaudin Hamiltonians appear in such equations. We will then review the KZ-BPZ relation in the $s\ell_2$ case; the reader is not advised to skip that section as the KZ-BPZ relation is presented in a form suitable for generalization to $s\ell_3$. In the $s\ell_3$ case, we will then find that the conjecture (\ref{omom}) holds only in the critical level limit $k\rar 3$.

\zeq\section{Gaudin Hamiltonians in conformal field theory}

We will review how the Gaudin Hamiltonians appear in Ward identities obeyed by correlation functions in conformal field theories with an $\widehat{s\ell_N}$ symmetry algebra. The Ward identities associated to the stress-energy tensor $T^J(z)$ lead to the KZ equations, which involve the ordinary Gaudin Hamiltonians. The Ward identities associated to the cubic field $W^J(z)$ involve higher Gaudin Hamiltonians. 

\subsection{Knizhnik--Zamolodchikov equations}

The affine Lie algebra $\widehat{s\ell_N}$ is an infinite-dimensional extension of the simple Lie algebra $s\ell_N$. The generators $t^a$ of $s\ell_N$, its structure constants $f^{ab}_c$, and its 
metric $\kappa^{ab}$ are defined by the relations
\bea
[t^a,t^b]=f^{ab}_c t^c\scs \kappa^{ab}\equiv \Tr t^at^b  \scs f^{ab}_c f^{cd}_b =2N \kappa^{ad}\ ,
\label{tfk}
\eea
where here and in the following the trace $\Tr$ is taken in the fundamental representation, so that our metric $\kappa^{ab}$ coincides with the renormalized Killing form of \cite{fms97}(13.13). The affine Lie algebra $\widehat{s\ell_N}$ can be formulated as the algebra of currents $J^a(z)$ with the operator product expansion
\bea
J^a(z)J^b(w) = -\frac{k \kappa^{ab}}{(z-w)^2} +f^{ab}_c \frac{J^c(w)}{z-w} + (J^aJ^b)(w) + {\cal O}(z-w)\ ,
\eea
where the parameter $k$ is called the level, and the normal-ordered product $(J^aJ^b)(w)$ is defined by the present formula. Conformal symmetry follows from the existence of a Virasoro algebra with central charge $c=\frac{k(N^2-1)}{k-N}$, generated by the Sugawara stress-energy tensor
\bea
T^J(z) \equiv -\frac{1}{2(k-N)} (J^aJ^a)(z)\ ,
\label{tz}
\eea
where $J^aJ^a$ is a shorthand for $\kappa_{ab}J^aJ^b$.
The identification of $T^J(z)$ with the generator of conformal transformations will be at the origin of the KZ equations. These equations are satisfied by any correlation function (\ref{omn}) of $n$ affine primary fields $\Phi^{j_i}(x_i|z_i)$ on the complex $z$-plane,
where the spins $j_i$ label representations of $s\ell_N$, the isospin variables $x_i$ label the states in a given representation, and the complex numbers $z_i$ are positions on the Euclidean two-dimensional spacetime. The affine primary fields are defined by their operator product expansions with the currents $J^a(z)$,
\bea
J^a(z) \Phi^{j}(x|w) = \frac{D^a\Phi^j(x|w)}{z-w} + {\cal O}(1)\ ,
\eea
where $D^a$ provides a realization of the representation of spin $j$ in terms of differential operators acting on the isospin variables $x$, so that $[D^a,D^b]=f^{ab}_c D^c$. We will keep this realization arbitrary, without committing to any particular choice of isospin variables. Let us however give an example of such a choice in the $s\ell_2$ case:
\bea
D^- = \pp{x} \scs D^3=x\pp{x} - j \scs D^+ = x^2\pp{x} -2jx\ .
\label{xbasis}
\eea
The KZ equations are now obtained by inserting $T^J(z)$ into the correlation function $\Omega_n$, and using the conformal Ward identity for $T^J(z)$ on the one hand, and the affine Ward identities for $(J^aJ^a)(z)$ on the other hand:
\bea
\la T^J(z) \prod_{i=1}^n \Phi^{j_i}(x_i|z_i)\ra &=& \sum_{i=1}^n \left(\frac{L^J_{0,(i)}}{(z-z_i)^2} + \frac{L^J_{-1,(i)}}{z-z_i}\right) \Omega_n \nn
\\
&=& -\frac{1}{2(k-N)}\sum_{i=1}^n \frac{D^a_{(i)}}{z-z_i}\sum_{\ell=1}^n\frac{D^a_{(\ell)}}{z-z_\ell}\ \Omega_n \ ,
\label{tward}
\eea
where the subscript $(i)$ in $D^a_{(i)}$ indicates that it acts on the isospin variables $x_i$, and
by definition $L^J_{p,(i)}$ is the $p$-th mode of $T^J(z)$ acting on $\Phi^{j_i}(x_i|z_i)$, according to 
\bea
L^J_p \Phi^{j}(x|z) \equiv \frac{1}{2\pi i}\oint_z dw\ (w-z)^{p+1} T^J(w)\Phi^j(x|z)\ .
\label{ljtj}
\eea
Calling $\Delta^J$ the eigenvalues of $L_0^J$, such that $L^J_{0,(i)}\Omega_n=\Delta^J_{j_i}\Omega_n$, we first deduce from eq. (\ref{tward}) the expression for $\Delta_j$ in terms of the quadratic Casimir $C_2(j)\equiv D^aD^a$ of the $s\ell_N$ representation with spin $j$,
\bea
\Delta^J_{j} \equiv -\frac{C_2(j)}{2(k-N)} .
\label{ljz}
\eea
Now $T^J(z)$ is assumed to generate conformal transformations, and in particular $L^J_{-1,(i)}\Omega_n=\ppl{z_i}\Omega_n $. 
(We define $\ppl{z_i}\equiv \left.\pp{z_i} \right|_{x_i}$ as a derivative at fixed isospin variables.)
Together with eq. (\ref{tward}), this implies the KZ equations \cite{kz84}
\bea
(k-N) \ppl{z_i} \Omega_n = -H_i \Omega_n \scs H_i \equiv \sum_{\ell\neq i} \frac{D^a_{(i)}D^a_{(\ell)}}{z_i-z_\ell}\ ,
\label{kz}
\eea
The $n$ commuting differential operators $H_i$ are called the Gaudin Hamiltonians. Through its dependence on $D^a_{(i)}$ and $D^a_{(\ell)}$, each one of the $n$ Hamiltonians involves all of the $n$ isospin variables $x_i$, which makes the problem of their simultaneous diagonalization difficult. This difficulty will be solved by Sklyanin's separation of variables, which replaces the isospins $x_i$ with new variables $y_i$, and combines the Gaudin eigenvalue equations into an essentially equivalent set of equations, each of which involves only one of the new variables.

\subsection{Ward identities for the cubic field \label{secwc}}

In addition to the quadratic invariant tensor $\kappa^{ab}=\Tr t^at^b$, it is possible to define the fully symmetric  cubic invariant tensor 
\bea
d^{abc} \equiv \Tr (t^at^bt^c + t^at^ct^b) \ .
\label{dabc}
\eea
This tensor vanishes in the case of $s\ell_2$, but not in the cases of $s\ell_{N\geq 3}$. It can then be used for constructing the invariant cubic field 
\bea
W^J(z) \equiv \frac16\rho\ d_{abc}(J^a(J^bJ^c))(z)\scs \rho\equiv \frac{i}{(k-N)^{\frac32}}\ .
\label{wz}
\eea
This generalizes the Sugawara construction (\ref{tz}), with however two substantial differences. 
First, while the field $T^J(z)$ is interpreted as the generator of conformal transformations, there is no such geometrical interpretation for $W^J(z)$. 
Second, while the field $T^J(z)$ obeys a Virasoro algebra, the field $W^J(z)$ does not obey the higher ${\cal W}_3$ algebra \cite{bs92}. In other words, while the Virasoro algebra can be realized as either a coset of $\widehat{s\ell_2}$ or a subalgebra of the enveloping algebra of $\widehat{s\ell_{N\geq 2}}$ (albeit with differing central charges), the ${\cal W}_3$ algebra is a coset of $\widehat{s\ell_3}$ but not a subalgebra of the enveloping algebra of $\widehat{s\ell_{N\geq 3}}$. 

In analogy with eq. (\ref{tward}) we now have
\bea
\la W^J(z) \prod_{i=1}^n \Phi^{j_i}(x_i|z_i)\ra &=& \sum_{i=1}^n \left(\frac{W^J_{0,(i)}}{(z-z_i)^3} + \frac{W^J_{-1,(i)}}{(z-z_i)^2} + \frac{W^J_{-2,(i)}}{z-z_i}\right) \Omega_n \nn
\\
&=& \frac16\rho\ d_{abc}\sum_{i=1}^n \frac{D^a_{(i)}}{z-z_i}\sum_{\ell=1}^n\frac{D^b_{(\ell)}}{z-z_\ell} \sum_{m=1}^n \frac{D^c_{(m)}}{z-z_m}\ \Omega_n \ ,
\label{wward}
\eea
where by definition $W^J_{p,(i)}$ is the $p$-th mode of $W^J(z)$ acting on $\Phi^{j_i}(x_i|z_i)$, according to 
\bea
W^J_p \Phi^{j}(x|z) \equiv \frac{1}{2\pi i}\oint_z dw\ (w-z)^{p+2} W^J(w)\Phi^j(x|z)\ .
\eea
Calling $q^J$ the eigenvalues of $W_0^J$, such that $W_{0,(i)}^J\Omega_n = q^J_{j_i}\Omega_n$, we first deduce from eq. (\ref{wward}) the expression for $q^J_j$ in terms of the cubic Casimir $C_3(j)\equiv d_{abc} (D^aD^bD^c+D^aD^cD^b)$ of the $s\ell_N$ representation with spin $j$,
\bea
q^J_j =\frac16 \rho\ C_3(j)\ .
\label{wzj}
\eea
We further deduce
\bea
W_{-1,(i)}^J \Omega_n &=& \frac12 \rho\ H'_i \Omega_n\ , 
\label{wuj}
\\
W_{-2,(i)}^J \Omega_n &=& \frac12 \rho\ H''_i \Omega_n\ ,
\label{wdj}
\eea
where the differential operators $H_i'$ and $H''_i$ are higher Gaudin Hamiltonians, whose explicit expressions in terms of $D^a_{(i)}$ can easily be derived from eq. (\ref{wward}).
But, in contrast to $L_{-1}^J$, the operators $W_{-1}^J$ and $W_{-2}^J$ are not interpreted as differential operators with respect to $z$. The equations (\ref{wuj}) and (\ref{wdj}), which generalize the KZ equations, are therefore not differential equations, and they will therefore not help us test our conjecture. Nevertheless, they will naturally appear in certain formulas. 

\zeq\section{Review of the $s\ell_2$ case}

In this section we will review the relation between the $s\ell_2$ KZ equations and BPZ equations. This was originally found by Feigin, Frenkel and Stoyanovsky \cite{sto00}, using Sklyanin's separation of variables for the $s\ell_2$ Gaudin model \cite{skl92c}. However, the original derivation relied on a particular choice of the isospin variables. This choice of isospin variables makes the result remarkably simple, but has no analog in the $s\ell_3$ case, as we will show. We will therefore reanalyze the $s\ell_2$ case, using whenever possible objects which do have analogs in the $s\ell_3$ or even $s\ell_N$ cases. We will present systematic derivations of their relevant properties, which will help clarify whether and how they can be generalized to the $s\ell_3$ case.

\subsection{Separation of variables for the $s\ell_2$ Gaudin model}

Let us consider a system of $n$ representations of $s\ell_2$ with spins $j_1\cdots j_n$. Consider the associated quantum variables $D^a_{(i)}$ such that $[D^a_{(i)},D^b_{(j)}]=\delta_{ij} f^{ab}_c D^c_{(i)}$ with $D^a_{(i)} D^a_{(i)} = C_2(j_i)$.
The system comes with parameters $z_1\cdots z_n$. 
Sklyanin's separation of variables for this system involves three ingredients: 
\begin{enumerate}
 \item A function $B(u)$ of an arbitrary variable $u$ (the spectral parameter), whose zeroes are the separated variables $y_i$, so that $B(y_i)=0$;
 \item Another function $A(u)$ such that $p_i = A(y_i)$ is the conjugate momenta to $y_i$;
 \item A kinematical identity, called the characteristic equation, which for any given $i$ relates $y_i$ and $p_i$.  
\end{enumerate}
We now briefly review the construction of these three objects in the $s\ell_2$ case. They are built from the $s\ell_2$ Lax matrix
\bea
I(u)\equiv -\sum_{i=1}^n \frac{t^a D^a_{(i)}}{u-z_i}\ ,
\label{iu}
\eea
whose matrix elements $I_\al^\be(u)$ obey the identity
\bea
(u-v) [I_\al^\g(u),I_\be^\e(v)] = \delta_\al^\e I_\be^\g(u)-\delta_\be^\g I_\al^\e(u) -\delta_\al^\e I_\be^\g(v)+\delta_\be^\g I_\al^\e(v)\ .
\label{uvii}
\eea
With the particular choice eq. (\ref{xbasis}) for the $s\ell_2$ isospin variable $x$, the $s\ell_2$ Lax matrix is explicitly
\bea
I(u) = -\left[\begin{array}{cc} \frac12\sum_{i=1}^n\frac{1}{u-z_i}\left(x_i\pp{x_i} -j_i\right) & \ \ \ \sum_{i=1}^n \frac{1}{u-z_i} \pp{x_i} \\ \sum_{i=1}^n\frac{1}{u-z_i} \left(x_i^2\pp{x_i} -2j_ix_i\right) & \ \ \ -\frac12 \sum_{i=1}^n \frac{1}{u-z_i}  \left(x_i\pp{x_i} -j_i\right) \end{array}\right]\ .
\eea
% In the derivation of this formula from the previous one, two minus signs will cancel: one from the manipulations
% of (u-z_i) factors, the other one from the ``propagation of the commutator'' between t^a and D^a.
Now choosing 
\bea
B(u)\equiv I_1^2(u) \scs A(u)\equiv I_1^1(u)\ ,
\eea
it is easy to check that
\bea
& & [B(u),B(v)]=0 \scs [A(u),A(v)]=0\ ,
\label{bbaa}
\\ {}
& & (u-v)[A(u),B(v)]=B(v)-B(u)\ .
\label{aubv}
\eea
These relations ensure that the operators $y_i$ defined as the zeroes of $B(u)$, and $p_i=A(y_i)$, do satisfy 
\bea
[y_i,y_j]=0  \scs \left[p_i ,y_j\right]=\delta_{ij} \scs [p_i,p_j]=0\ .
\eea
In particular, $\left[p_i ,B(v)\right]=\frac{B(v)}{y_i-v}$ agrees with $B(v)\propto \frac{\prod_i (v-y_i)}{\prod_j (v-z_j)}$.
There is however a problem of operator ordering in the expressions $A(y_i)$ and $B(y_i)$, because the separated variables $y_i$ are operators. This problem is dealt with in reference \cite{skl92c}. We will ignore it in the  forthcoming heuristic derivation of the characteristic equation.
Let us start with $\det\left( A(y_i)\id - I(y_i) \right)=0$, where $\id$ is the identity matrix. (The determinant of a matrix whose first line vanishes is zero.) This implies $p_i^2  - \frac12 (I_\al^\be I_\be^\al)(y_i)=0$. This characteristic equation can easily be rewritten as
\bea
\boxed{ p_i^2 -\frac12\sum_\ell \frac{C_2(j_\ell)}{(y_i-z_\ell)^2} -\sum_\ell\frac{1}{y_i-z_\ell}H_\ell = 0\ ,
}\label{cheq}
\eea
where $H_\ell$ is of course a Gaudin Hamiltonian (\ref{kz}), and $C_2(j)$ is the quadratic Casimir of a spin-$j$ representation.

\paragraph{Functional space interpretation.}

We now wish to consider the quantum variables $D^a_{(i)}$ as differential operators acting on functions $\Psi(\{x_i\})$ of isospin variables $x_i$. (An example of such a realization was given in eq. (\ref{xbasis}).) Similarly, the separated variables $y_\ell$ and their associated momenta $p_\ell$ may act on functions $\tilde{\Psi}(\{y_\ell\})$, in particular $p_\ell \tilde{\Psi} = \pp{y_\ell} \tilde{\Psi}$. The separation of variables $\{x_i\} \rar \{y_\ell\},\ev$ (where the extra variable $\ev$ will be defined shortly) is then intepreted as an integral transformation ${\cal K}$ such that
\bea
\Psi(\{x_i\}) = {\cal K}\ \tilde{\Psi}(\{y_\ell\},\ev) =  \int d\ev\ \int \prod_\ell dy_\ell\ K(\{x_i\}|\{y_\ell\},\ev)\ \tilde{\Psi}(\{y_\ell\},\ev)\ ,
\label{pkp}
\eea
where the kernel $K$ is characterized as a common eigenvector of the commuting operators $B(u)$
\bea
\left(B(u) - \ev \frac{\prod_\ell (u-y_\ell)}{\prod_i (u-z_i)} \right) K(\{x_i\}|\{y_\ell\},\ev) = 0\ .
\label{bk}
\eea
The simultaneous diagonalization of the Gaudin Hamiltonians $H_j$, namely the set of equations $(H_\ell-E_\ell)\Psi =0$, can now be reformulated using the characteristic equation (\ref{cheq}), which implies
\bea
\left(\ppd{y_i} -\frac12\sum_\ell \frac{C_2(j_\ell)}{(y_i-z_\ell)^2} -\sum_\ell\frac{E_\ell}{y_i-z_\ell}\right) \tilde{\Psi} = 0\ ,
\eea
The solutions of this equation can be found in factorized form $\tilde{\Psi}=\prod_i \tilde{\psi}(y_i)$. This justifies the name ``separation of variables'' attributed to the change of variables $x_i\rar y_i$.

\paragraph{Some remarks.}

Finding the kernel $K$ by 
the simultaneous diagonalization of the operators $B(u)$ is easy in the $s\ell_2$ case because $B(u)=I_1^2(u)$ is a sum of $n$ commuting operators, so that we have $K(\{x_i\}|\{y_\ell\},\ev)=\prod_{i=1}^n k_i(x_i|\{y_\ell\},\ev)$ where the the equation on $k_i$ is obtained from eq. (\ref{bk}) in the limit $u\rar z_i$:
\bea
\left( (t^a)_1^2 D^a_{(i)} + \mu_i\right) k_i(x_i|\{y_\ell\},\ev) = 0 \scs \mu_i\equiv \ev \frac{\prod_\ell (z_i-y_\ell)}{\prod_{j\neq i} (z_i-z_j)}\ .
\label{mudef}
\eea
For example, if the isospin variables are chosen as in eq. (\ref{xbasis}), then we find $k_i=e^{-\mu_i x_i}$. This suggests that we could use other isospin variables $\hat{\mu}_i$ such that $D^a_{(i)} (t^a)_1^2 = -\hat{\mu}_i$, then we would find $k_i\propto\delta(\hat{\mu}_i-\mu_i)$, so that we could explicitly perform the integrals in eq. (\ref{pkp}). This would lead to $\Psi(\{\hat{\mu}_i\}) \propto \Psi(\{y_\ell\},\ev)$ with simple proportionality factors, as the change of variables $\{\hat{\mu}_i\} \rar (\{y_\ell\},\ev)$ would now be local and described by the functions $\mu_i(\{y_\ell\},\ev)$. More generally, for any choice of isospin variables, the kernel $K$ will be of the type
\bea
K(\{x_i\}|\{y_\ell\},\ev|\{z_j\}) = \prod_{i=1}^n \left. k_i\left(x_i\right|\{\mu_j\}\right)\ , 
\label{kki}
\eea
where $\mu_j(\{y_\ell\},U|\{z_j\})$ is defined in eq. (\ref{mudef}), and we made the $z_j$-dependence explicit. Thus, in the $s\ell_2$ case, the kernel $K$ can be determined explicitly, and this is because the operator $B(u)$ is a linear function of the Lax matrix $I(u)$.

Let us finally be more precise about the number of variables $y_\ell$. They are defined as the zeroes of a rational function $B(u)$ which, barring extra constraints, has $n$ poles and degree $-1$. Therefore we must have $n-1$ such variables, and the $n$th variable $\ev$ is the eigenvalue of $-(t^a)_1^2\sum_{i=1}^n  D^a_{(i)} $. In conformal field theory applications, we however impose the extra constraint $\sum_{i=1}^n D^a_{(i)}=0$, so that $B(u)$ has degree $-2$. This yields $n-2$ variables $\{y_\ell\}_{\ell=1\cdots n-2}$, and $\ev$ is the eigenvalue of $-(t^a)_1^2\sum_{i=1}^n z_i  D^a_{(i)} $.

\subsection{The $s\ell_2$ Knizhnik--Zamolodchikov equations in Sklyanin variables}

We just saw that Sklyanin's separation of variables is useful tool for simultaneously diagonalizing the $s\ell_2$ Gaudin Hamiltonians. This problem is closely related to the problem of solving the KZ equations (\ref{kz}), which are obtained by replacing the eigenvalues of the Gaudin Hamiltonians $H_i$ with $-(k-2)\ppl{z_i} $. This suggests that it may be interesting to rewrite the KZ equations in terms of Sklyanin's variables. To do this, we will use the characteritic equation (\ref{cheq}) which such variables obey, and apply it to ${\cal K}^{-1}\Omega_n$, which is a function of $\{y_i\}$, so that $p_i{\cal K}^{-1}\Omega_n =\pp{y_i} {\cal K}^{-1}\Omega_n$. While itself just a kinematical identity, the characteristic equation then allows us to reorganize the KZ equations as
\bea
\boxed{\left(\frac{1}{k-2}\ppd{y}  +\sum_{\ell=1}^n \frac{1}{y-z_\ell} {\cal K}^{-1}\ppl{z_\ell} {\cal K}+ \sum_{\ell=1}^n \frac{\Delta^J_{j_\ell}}{(y-z_\ell)^2}\right) {\cal K}^{-1}\ \Omega_n = 0\ , }
\label{chkz}
\eea 
where we drop the index from $y_i$, and we use $\Delta^J_j=-\frac{C_2(j)}{2(k-2)}$ from eq. (\ref{ljz}). We still have to perform the change of variables on the
$z_\ell$-derivatives at fixed isospins, i.e. to rewrite $ {\cal K}^{-1}\ppl{z_\ell} {\cal K}$ in terms of $\pp{z_\ell}\equiv \left.\pp{z_\ell} \right|_{y_a}$. This is rather easy because of the particular form of the kernel (\ref{kki}), where the dependences on $\{y_a\},\ev$ and $\{z_\ell\}$ are channeled through the particular functions $\{\mu_i\}$. This implies that the integral transformation (\ref{pkp}) just adds first-order differential operators $\pp{y_a} ,\pp{\ev} $ to $\ppl{z_\ell}$, so that 
\bea
 {\cal K}^{-1}\ppl{z_\ell} {\cal K} = \pp{z_\ell} + \sum_a \left.\frac{\p y_a}{\p z_\ell}\right|_{\mu_i} \pp{y_a} + \left.\frac{\p \ev}{\p z_\ell}\right|_{\mu_i} \pp{\ev} \ .
\label{ppz}
\eea
Denoting $\{y_a\}=\{y,\{y_b\}\}$, we obtain the KZ equations in Sklyanin variables,
\begin{multline}
\left(\frac{1}{k-2}\ppd{y} +\sum_{\ell=1}^n \frac{1}{y-z_\ell} \left(\pp{z_\ell} +\pp{y} \right) +\sum_b \frac{1}{y-y_b}\left(\pp{y_b} -\pp{y} \right) \right.
\\ \left. + \sum_{\ell=1}^n \frac{\Delta^J_{j_\ell}}{(y-z_\ell)^2} \right){\cal K}^{-1}\ \Omega_n = 0\ .
\label{kzsep}
\end{multline}
In this equation the variables are no longer separated, as the variables $y_b$ appear in addition to $y$.

\subsection{Comparison with Virasoro null-vector equations}

In the previous subsection, we have studied the KZ equations in a CFT with an $\widehat{s\ell_2}$ symmetry algebra at level $k$. We will now compare them with null-vector equations in a CFT with a Virasoro symmetry algebra at central charge $c=1+6(b+b^{-1})^2$ where $b^2\equiv \frac{1}{k-2}$. This is the Virasoro algebra which would be obtained from our $\widehat{s\ell_2}$ algebra by quantum Hamiltonian reduction (see for instance \cite{bs92}), although that reduction does not explain the relation between differential equations which we are about to review. 

The Virasoro algebra can be formulated in terms of the stress-energy tensor $T(z)$, which obeys
\bea
T(z)T(w)=\frac{\frac12 c}{(z-w)^4} + \frac{2T(w)}{(z-w)^2} + \frac{\p T(w)}{z-w} + {\cal O}(1)\ .
\eea
Primary fields $V_\al(w)$ of momentum $\al$ and conformal dimention $\Delta_\al = \al(b+b^{-1}-\al)$ are defined by
\bea
T(z) V_\al(w) = \frac{\Delta_\al V_\al(w)}{(z-w)^2} + \frac{\p V_\al(w)}{z-w} + {\cal O}(1)\ .
\label{tva}
\eea
This definition does not distinguish the primary fields $V_\al$ and $V_{b+b^{-1}-\al}$, which have the same conformal dimension. These fields are therefore assumed to be proportional, with a proportionality constant called the reflection coefficient. This $\Z_2$ symmetry can be understood as the action of the Weyl group of $s\ell_2$ on the space of the momenta $\al$. 

The Virasoro representation generated by the degenerate field $V_{-\frac{1}{2b}}$ is known to have a null-vector at level two Namely, $(L_{-2}+b^2L_{-1}^2) V_{-\frac{1}{2b}}=0$, where the modes $L_p$ are defined as in eq. (\ref{ljtj}). This implies that correlation functions involving such a degenerate field obey the Belavin--Polyakov--Zamolodchikov equation \cite{bpz84}
\bea
\left[b^2\ppd{y} + \sum_{i=1}^n \frac{1}{y-z_i}\pp{z_i} + \sum_{i=1}^n \frac{\Delta_{\al_i}}{(y-z_i)^2} \right]
\ \la V_{-\frac{1}{2b}}(y) \prod_{i=1}^n V_{\al_i}(z_i) \ra = 0\ .
\label{bpz}
\eea
Curiously, this equation is formally identical to the variable-separated KZ equation (\ref{chkz}). The meaning of this formal similarity is not clear to us. The KZ equations in Sklyanin variables (\ref{kzsep}) actually involve $n-2$ variables $y_1\cdots y_{n-2}$, therefore we should rather consider correlation functions of the type
\bea
\tilde{\Omega}_n \equiv \la \prod_{a=1}^{n-2} V_{-\frac{1}{2b}}(y_a) \prod_{i=1}^n V_{\al_i}(z_i) \ra \ . 
\eea
We then expect such correlation functions to be related to $\Omega_n$ (\ref{omn}) as in equation (\ref{omom}). 
That equation means that the twisted BPZ equations satisfied by $\Theta_n \tilde{\Omega}_n$ are identical to the KZ equations in Sklyanin variables (\ref{kzsep}). This can indeed be checked by explicit calculation, provided we correctly specify the function $\Theta_n$ as well as the relation between $s\ell_2$ spins $j_i$ and Virasoro momenta $\al_i$. Requiring that the $\al-j$ relation is compatible with the respective Weyl symmetries $j\rar -j-1$ and $\al\rar b+b^{-1}-\al$, and that conformal dimensions $\Delta^J_j=-\frac{j(j+1)}{k-2}$ eq. (\ref{ljz}) and $\Delta_\al$ are related by a constant shift, determines the relation
\bea
\al = b(j+1) +\frac{1}{2b} \scs \Delta_\al = \Delta^J_j +\frac12 +\frac{1}{4b^2}\ .
\eea
We still have to specify the values of the parameters $\lambda,\mu,\nu$ in the ansatz (\ref{thn}) for the function $\Theta_n$. We could determine these values by requiring the twisted BPZ equations to agree with eq. (\ref{kzsep}), and we would find
\bea
\lambda = \frac{1}{2b^2} \scs \mu =-\frac{1}{2b^2} \scs \nu=\frac{1}{2b^2} \ .
\label{lmnv}
\eea
There are simple concurring arguments for the values of $\lambda$ and $\nu$.
First, the value of $\lambda$ is determined by the requirement of continuity of $\Theta_n \tilde{\Omega}_n$ at $y_a=y_b$. This requirement plays an important role in the boundary $\Hp$ model \cite{hr06}. 
Second, the value of $\nu$ follows from checking equation (\ref{omom}) in the simplest case $n=2$, when there are no $y_a$ variables and no BPZ equations. 

Let us now comment on this twist function $\Theta_n$ and its relation to free field correlation functions. In this paragraph we will consider full correlation functions with dependences on both holomorphic and antiholomorphic variables, and the full twist factor which is thus $|\Theta_n|^2$. 
With the above values (\ref{lmnv}) for $\lambda,\mu,\nu$, we observe that the inverse twist factor $|\Theta_n|^{-2}$ coincides with the free field correlation function formally obtained from $\tilde{\Omega}_n$ by taking the fields $V_{\al_i}(z_i)$ to have momenta $\al_i=\frac{1}{2b}$ instead of $\al_i=b(j_i+1)+\frac{1}{2b}$. This means 
\bea
|\Theta_n|^{-2}=\la \prod_{a=1}^{n-2} V_{-\frac{1}{2b}}(y_a) \prod_{i=1}^n V_{\frac{1}{2b}}(z_i) \ra^{free}\ .
\label{thav}
\eea 
This interpretation of $\Theta_n$ plays a role in a recent proof of the FZZ conjecture \cite{hs08}, see also \cite{gl07}. For now, let us explain the origin of this observed relation by studying the $b\rar 0$ limit of the $\Hp$-Liouville relation. This relation can be written as $\Omega_n \sim {\cal K}{\cal \bar{K}} |\Theta_n|^2\tilde{\Omega}_n$\ , whose factors we now analyze:
\begin{itemize}
\item
The Liouville correlation function $\tilde{\Omega}_n$ reduces to $\la \prod_{a=1}^{n-2} V_{-\frac{1}{2b}}(y_a) \prod_{i=1}^n V_{\frac{1}{2b}}(z_i) \ra$ as $b\rar 0$. And it turns out that this coincides with a free field correlation function, because the momentum conservation condition is obeyed. Namely, the sum of the momenta is $(n-2)\times -\frac{1}{2b} + n\times \frac{1}{2b} = \frac{1}{b}$ which coincides with the dominant term in the Liouville background charge $\frac{1}{b}+b$. Therefore, according to standard path-integral reasoning in Liouville theory \cite{zz95}, we have $\tilde{\Omega}_n \underset{b\rar 0}{\sim}R_n \la \prod_{a=1}^{n-2} V_{-\frac{1}{2b}}(y_a) \prod_{i=1}^n V_{\frac{1}{2b}}(z_i) \ra^{free} =R_n \left|\frac{\prod (y_a-z_i)}{\prod (y_a-y_b)\prod(z_i-z_j)}\right|^{\frac{1}{b^2}} $ where $R_n$ is $b$-independent. 
\item
The $\Hp$ correlation function $\Omega_n$ is expected to have a finite ``minisuperspace'' limit \cite{tes97b} as $b\rar 0$ which is equivalent to $k\rar \infty$ where $k$ is the level.
\item
The separation of variables ${\cal K}$ is $b$-independent by definition.
\item
So the twist factor $|\Theta_n|^2$ must absorb the $b\rar 0$ divergence of the Liouville correlation function $\tilde{\Omega}_n$, which implies the relation (\ref{thav}) and the values (\ref{lmnv}) for the parameters $\lambda,\mu,\nu$. (This reasoning does not exclude the presence of extra terms in $\lambda,\mu,\nu$ which would be finite in the $b\rar 0$ limit.) 
\end{itemize}

This concludes our reminder of the KZ-BPZ relation in the $s\ell_2$ case. In the next section we will analyze the $s\ell_3$ KZ equations along the same lines.

\zeq\section{The $s\ell_3$ case}

\subsection{Separation of variables for the $s\ell_3$ Gaudin model}

To the best of our knowledge, the full quantum separation of variables for the $s\ell_3$ Gaudin model has not been derived yet. By the full separation of variables we mean the determination of $A(u)$, $B(u)$ and a characteristic equation, like in the $s\ell_2$ case.\footnote{A different approach was proposed in \cite{mstv06}, which consists in trying to use the $s\ell_2$ separation of variables in the $s\ell_3$ case. This approach requires a particular choice of isospin variables. The results are complicated.}
Sklyanin did however derive the full separation of variables for the classical $s\ell_3$ Gaudin model \cite{skl92}. In order to derive the quantum version, we will use Sklyanin's separation of variables for models with an $s\ell_3$ Yangian symmetry \cite{skl92b}, see also \cite{smi01} for a generalization to $s\ell_N$. This Yangian symmetry is present in the Gaudin model, which will allow us to derive its quantum characteristic equation from the Yangian's.

\paragraph{$s\ell_3$ Yangian symmetry.}

As in the $s\ell_2$ case, the variables of the $s\ell_N$ Gaudin model can be combined into an $s\ell_N$ Lax matrix $I(u)$ (\ref{iu}) obeying the relation (\ref{uvii}). It is however possible to combine the variables into another $s\ell_N$ matrix, which depends on an extra parameter $\eta$,
\bea
Y(u) &\equiv& \left(\id - \frac{\eta}{u-z_1}t^a D^a_{(1)}\right) \left(\id - \frac{\eta}{u-z_2}t^a D^a_{(2)}\right)\cdots  \left(\id - \frac{\eta}{u-z_n}t^a D^a_{(n)}\right)
\label{hu}
\\
&=& \id + \eta I(u) + \frac12 \eta^2 :I^2:(u) +\frac16 \eta^3 :I^3:(u) + \cdots\ ,
\label{hexp}
\eea
where the definition of the normal ordering in $:I^2:(u)$ and $:I^3:(u)$ follows from the chosen ordering of the factors of $Y(u)$. This object can be shown to obey the Yangian algebra
\bea
(u-v)Y_\al^\g(u)Y_\be^\e(v) + \eta Y_\al^\e(u)Y_\be^\g(v) = (u-v)Y_\be^\e(v) Y_\al^\g(u) +\eta Y_\al^\e(v)Y_\be^\g(u)\ .
\eea
Sklyanin's separated variables $y_\ell$ for the Yangian \cite{skl92b} are defined as the zeroes of a function 
\begin{multline}
B^Y(u) = Y_3^2(u)Y_2^1(u)Y_3^2(u-\eta) -Y_3^2(u)Y_3^1(u)Y_2^2(u-\eta) \\ +Y_3^1(u)Y_3^2(u)Y_1^1(u-\eta)-Y_3^1(u)Y_1^2(u)Y_3^1(u-\eta)\ ,
\end{multline}
while the conjugate variables  are given by $X_i = A^Y(y_i)$ where
\bea
A^Y(u) = Y_1^1(u) - Y_3^2(u-\eta)^{-1} Y_3^1(u-\eta) Y_1^2(u)\ ,
\eea
Let us point out that interesting structural insight into these formulas for $A^Y(u)$ and $B^Y(u)$ was obtained in \cite{cf07}, based on general properties of matrices with non-commuting elements. 
The functions $A^Y(u)$ and $B^Y(u)$ obey the commutation relations
\begin{multline}
[A^Y(u),A^Y(v)]=0 \scs [B^Y(u),B^Y(v)]=0 \scs
\frac{u-v}{\eta} [A^Y(u),B^Y(v)]
\\
=B^Y(u)A^Y(v)\ Y_3^2(u-\eta)^{-1} Y_3^2(u)^{-1} Y_3^2(v-\eta)Y_3^2(v) -B^Y(v) A^Y(u) \ ,
\label{ayby}
\end{multline}
so that 
\bea
[y_i,y_j]=0 \scs [X_i,y_j]=-\eta \delta_{ij}X_i \scs [X_i,X_j]=0\ .
\eea
The quantum characteristic equation is then
\bea
X_i^3 - X_i^2 t_1(y_i) +X_i t_2(y_i-\eta) - d(y_i-2\eta) =0\ ,
\label{ych}
\eea
with the invariant operators $t_1(u)$, $t_2(u)$ and $d(u)$ defined as \cite{skl92b} 
\bea
t_1(u)=\Tr Y(u) \scs t_2(u)=\Tr \tilde{Y}(u) \scs d(u)\delta_\al^\g = Y_\al^\be(u){} \tilde{Y}_\be^\g(u+\eta)\ ,
\eea
where the matrix $\tilde{Y}$ is constructed by transposing the quantum comatrix of $Y$. For instance, $\tilde{Y}_3^2(u)=-Y_3^2(u)Y_1^1(u+\eta)+Y_3^1(u)Y_1^2(u+\eta)$, where the $\eta$-shifts are the manifestation of the quantum character of the comatrix whose $_2^3$ matrix element we just wrote. Operator ordering issues in expressions like $t_2(y_i-\eta)$ are resolved by inserting the operator $y_i$ from the left. 

\paragraph{From the Yangian to the Gaudin model.}

We will now construct objects $A(u)$, $B(u)$ and a quantum characteristic equation for the $s\ell_3$ Gaudin model. Such $\eta$-independent functions of the matrix $I(u)$ will be obtained by
expanding the corresponding objects for the $s\ell_3$ Yangian algebra in powers of $\eta$. We find
\bea
A^Y(u) =& 1 - \eta A(u) + {\cal O}(\eta^2)\ , &\ \ A(u) = -I_1^1 + \frac{I_3^1 I_1^2}{I_3^2} \ ,
\label{ayau}
\\
B^Y(u) =& \eta^3 B(u) + {\cal O}(\eta^4)\ , &\ \  B(u) = I_2^1I_3^2I_3^2 - I_3^2I_3^1I_2^2 + I_3^1I_3^2I_1^1 - I_1^2I_3^1I_3^1\ ,
\label{bybu}
\eea
where we omitted the spectral parameter $u$ in $I_\al^\be(u)$, and we point out that our formula for $A(u)$ is free of ordering ambiguities because $I_3^2(u)$ commutes with both $I_1^2(u)$ and $I_3^1(u)$. The commutation relations (\ref{ayby}) for $A^Y(u)$ and $B^Y(u)$ imply the analogous relations
\bea
& & [A(u),A(v)]=0\scs [B(u),B(v)]=0\ ,
 \\
 & & (u-v) [A(u),B(v)]= B(v) -B(u) \frac{I_3^2(v) I_3^2(v)}{I_3^2(u) I_3^2(u)} \ ,
\eea 
which may be compared to the corresponding relations in the $s\ell_2$ case eq. (\ref{aubv}).

Let us rewrite the characteristic equation (\ref{ych}) as:
\begin{multline}
(X_i-1)^3 -(X_i-1)^2 \left[t_1(y_i)-3\right]+(X_i-1)\left[t_2(y_i-\eta)-2t_1(y_i)+3\right] 
\\
+ \left[1-t_1(y_i)+t_2(y_i-\eta)-d(y_i-2\eta)\right] = 0 \ .
\label{och}
\end{multline}
The leading behaviour of this equation as $\eta\rar 0$ will turn out to be ${\cal O}(\eta^3)$. To compute this behaviour, we of course need to compute the behaviours of $X_i$ and $y_i$ as $\eta \rar 0$. It turns out that we only need the ${\cal O}(\eta)$ behaviour of $X_i$. We therefore define the variable $p_i$ by $X_i =1-\eta p_i +{\cal O}(\eta^2)$. As for $y_i$ we only need need the leading ${\cal O}(1)$ behaviour. To this leading order, the zeroes of $B^Y(u)$ coincide with those of $B(u)$, so that we do not need distinct notations and call them all $y_i$. The most complicated part of the calculation however does not involve such subtleties, but rather deals with the last term in eq. (\ref{och}),
\begin{multline}
 1-t_1(u)+t_2(u-\eta)-d(u-2\eta) 
 \\
 \begin{array}{l}
 = (1-Y_1^1(u-2\eta))(Y_3^3(u-\eta)-1)Y_2^2(u) +(Y_1^1(u-\eta)-Y_1^1(u-2\eta))Y_2^2(u) 
 \\ \ \ 
 +(Y_1^1(u-2\eta)-1)Y_3^2(u-\eta)Y_2^3(u) +(Y_1^1(u-\eta)-1)Y_3^3(u) +(1-Y_1^1(u))
  \\ \ \ 
  + (Y_3^3(u-2\eta)-1)Y_1^2(u-\eta)Y_2^1(u) -Y_3^2(u-2\eta)Y_1^3(u-\eta)Y_2^1(u) 
  \\ \ \ 
  -Y_1^2(u-2\eta)Y_3^1(u-\eta)Y_2^3(u) +Y_1^3(u-2\eta)Y_3^1(u-\eta)Y_2^2(u) -Y_1^3(u-\eta)Y_3^1(u)
   \end{array}
 \\ 
 \begin{array}{l}
 = \eta^3\left[ -I_1^1I_3^3I_2^2+I_1^1I_3^2I_2^3+I_3^3I_1^2I_2^1+I_2^2I_1^3I_3^1-I_3^2I_1^3I_2^1-I_1^2I_3^1I_2^3 
\right. \\ \left.
 \hspace{1cm} +I_1^1(I')_3^3-I_1^1(I')_1^1-I_1^3(I')_3^1 -(I')_1^3I_3^1 -(I'')_1^1\right] + {\cal O}(\eta^4)\ ,  
 \end{array}
\end{multline}
where we omitted the spectral parameter $u$ in $I_\al^\be(u)$, and used the $s\ell_3$-defining relation $I_1^1 + I_2^2 + I_3^3 = 0$.
We then obtain the following quantum characteristic equation of the $s\ell_3$ Gaudin model:
\bea
\boxed{
p_i^3 - p_i \cdot \frac12 (I_\al^\be I_\be^\al)(y_i) +\frac14 (I_\al^\be I_\be^\al)'(y_i)+\frac16\left(I_\al^\be I_\be^\g I_\g^\al + I_\be^\al I_\g^\be I_\al^\g\right)(y_i) = 0 \ . }
\label{qqq}
\eea
Notice that the particular cubic invariant which appears in this formula is related to the fully symmetric invariant tensor $d_{abc}$ eq. (\ref{dabc}). Using the definition (\ref{iu}) of $I(u)$, we indeed have
\bea
\left(I_\al^\be I_\be^\g I_\g^\al + I_\be^\al I_\g^\be I_\al^\g\right)(u) = -d_{abc} \sum_{i=1}^n \frac{D^a_{(i)}}{u-z_i}\sum_{\ell=1}^n\frac{D^b_{(\ell)}}{u-z_\ell} \sum_{m=1}^n \frac{D^c_{(m)}}{u-z_m}\ .
\eea
This could further be expressed in terms of the higher Gaudin Hamiltonians of Section \ref{secwc}, so that the characteristic equation could help simultaneously diagonalize these Hamiltonians.

\paragraph{Some remarks.} 

Like in the $s\ell_2$ case, Sklyanin's change of variables can be interpreted as an integral transformation ${\cal K}$ (\ref{pkp}) acting on a functional space. The kernel $K$ of ${\cal K}$ now obeys
\bea
\left(B(u) - \ev \frac{\prod_\ell (u-y_\ell)}{\prod_i (u-z_i)^3} \right) K(\{x_i\}|\{y_\ell\},\ev) = 0\ .
\eea
However, the simultaneous diagonalization of the commuting operators $B(u)$ is now a difficult problem, as $B(u)$ is now cubic and not linear in $I(u)$, and thus no longer a sum of $n$ commuting operators. Therefore, the kernel $K$ is no longer of the form (\ref{kki}). Certainly, no choice of isospin variables exists such that the kernel $K$ has a simple expression.
Another difference with the $s\ell_2$ case is the counting of variables: generic functions of the $s\ell_3$ isospin coordinates $x_i$ should correspond to functions of not only $y_i$ and $\ev$, but also of two extra variables. These extra variables are necessary for the transformation ${\cal K}$ to be invertible. We will neglect this issue\footnote{A construction of the extra variables seems to be available in the article \cite{ahh96}.}, as well as the issue of precisely defining the relevant functional spaces, and we will assume ${\cal K}$ to be invertible.

Let us finally determine the number of separated variables $y_i$ -- that is, the number of zeroes of $B(u)$. Barring extra constraints, this is of course $3n-3$. In conformal field theory applications, we however impose the extra constraints $\sum_{i=1}^n D^a_{(i)}=0$, so that $I(u)$ has degree $-2$. This does not immediately imply that $B(u)$ (eq. (\ref{bybu})), which is cubic in $I(u)$, has degree $-6$, because $\sum_{i=1}^n D^a_{(i)}=0$ only holds when directly applied to a physical correlation function, and
the matrix elements of $I(u)$ generically do not commute with each other. Rather, the degree of $B(u)$ depends on its precise form and should be evaluated by explicit calculation. We find that each one of the four terms of $B(u)$ has degree $-5$, while $B(u)$ itself has degree $-6$. This means that there are $3n-6$ separated variables. Therefore, as in the $s\ell_2$ case, the number of separated variables vanishes for $n=2$.

\subsection{The $s\ell_3$ Knizhnik--Zamolodchikov equations in Sklyanin variables}

Let us consider a conformal field theory with an $\widehat{s\ell_3}$ symmetry algebra. The Ward identities consist in the $n$ KZ differential equations (\ref{kz}), plus $2n$ extra non-differential relations (\ref{wuj}) and (\ref{wdj}), which express $W_{-1,(i)}^J$ and $W_{-2,(i)}^J$ in terms of differential operators acting on isospin variables. Let us reorganize all these relations by injecting them into the characteristic equation of the quantum $s\ell_3$ Gaudin model (\ref{qqq}). The result is schematically of the form 
\bea
\left[\ppt{y} +(k-3) \pp{y}\cdot T^J(y) -\frac12(k-3) \p T^J(y) -\frac{1}{\rho} W^J(y)\right] {\cal K}^{-1} \Omega_n =0\ ,
\label{schkz}
\eea
where the constant $\rho$ was defined in eq. (\ref{wz}). Explicitly,
\begin{multline}
\left[ \ppt{y} + (k-3)\pp{y}\cdot \sum_{i=1}^n\left(\frac{1}{y-z_i}{\cal K}^{-1}\ppl{z_i}{\cal K} + \frac{\Delta^J_{j_i}}{(y-z_i)^2} \right)\right.
\\ +\frac12(k-3) \sum_{i=1}^n \left(\frac{1}{(y-z_i)^2}{\cal K}^{-1}\ppl{z_i} {\cal K}+ \frac{2 \Delta^J_{j_i}}{(y-z_i)^3} \right)
\\ \left. -\frac{1}{\rho} \sum_{i=1}^n \left(\frac{{\cal K}^{-1}W_{-2,(i)}^J{\cal K}}{y-z_i} + \frac{{\cal K}^{-1}W^J_{-1,(i)}{\cal K}}{(y-z_i)^2} + \frac{q^J_{j_i}}{(y-z_i)^3} \right)\right] {\cal K}^{-1} \Omega_n = 0\ , 
\label{sepkz}
\end{multline}
where $\Omega_n$ is still an $n$-point function of the type (\ref{omn}).

In this equation, the terms involving $W_{-1,(i)}^J$ and $W_{-2,(i)}^J$ refer to correlation functions involving descendents of the primary fields $\Phi^{j}(\mu|z)$. We have little control over such non-differential terms, and we would like to ignore them in the following. This could be done by considering appropriate linear combinations of our $3n-6$ equations. (Remember that the variable $y$ spans the $3n-6$ separated variables $\{y_a\}$). We will for simplicity adopt the alternative approach of working modulo the unwanted terms. Let us make this precise by defining  the space ${\cal D}_S$ of differential operators in $y_a,z_i$ (including functions of $y_a,z_i$) which are symmetric under permutations of $\{y_1,y_2\cdots y_{3n-6}\}$. For any choice $\{y_a\} = \{y,y_b\}$ of a distinguished variable $y$ we further define 
\bea
{\cal F}_2(y) \equiv \sum_{i=1}^n \frac{1}{y-z_i} {\cal D}_S  +\sum_{i=1}^n \frac{1}{(y-z_i)^2} {\cal D}_S\ .
\label{fty}
\eea
By a simple counting of variables it can be realized that any differential operator which is symmetric under permuations of $\{y_b\}$ does belong to ${\cal F}_3(y) \equiv \sum_{i=1}^n \frac{1}{y-z_i} {\cal D}_S  +\sum_{i=1}^n \frac{1}{(y-z_i)^2} {\cal D}_S +\sum_{i=1}^n \frac{1}{(y-z_i)^3} {\cal D}_S$. But it does not always belong to ${\cal F}_2(y)$, so we can define a nontrivial equivalence $\sim$ as the equality modulo ${\cal F}_2(y)$. Thus, equation (\ref{sepkz}) simplifies to
\begin{multline}
\hspace{-4mm}\boxed{
\left[ \ppt{y} +\pp{y}\cdot \sum_{i=1}^n\frac{k-3}{y-z_i}{\cal K}^{-1}\ppl{z_i}{\cal K} + \sum_{i=1}^n \frac{(k-3)\Delta_{j_i}^J}{(y-z_i)^2}\pp{y} - \sum_{i=1}^n \frac{\frac{1}{\rho } q^J_{j_i} + (k-3) \Delta^J_{j_i}}{(y-z_i)^3} \right]\!
{\cal K}^{-1}\Omega_n \sim 0
} \\
\label{sepmod}
\end{multline} 
Having thus eliminated $W_{-1,(i)}^J$ and $W_{-2,(i)}^J$, we are left with operators $\ppl{z_i} $, which we recall are $z_i$-derivatives at fixed isospin variables. We expect ${\cal K}^{-1}\ppl{z_i}{\cal K}$ to be a combination of the operators $\pp{z_i}$, $\pp{y_a}$ and $\pp{U}$, although we do not know how to compute it. And it is not clear whether ${\cal K}^{-1}\ppl{z_i}{\cal K}$ is a first-order differential operator, as happened in the $s\ell_2$ case (see eq. (\ref{ppz})). Nevertheless, we do know that ${\cal K}^{-1}\ppl{z_i}{\cal K}$ is independent from the level $k$, which is a parameter of our conformal field theory but neither of the Gaudin model nor of its separation of variables. Therefore, we will still be able to extract useful information from eq. (\ref{sepmod}), a sum of terms with various power-like dependences on $(k-3)$, by considering all terms which are not linear in $(k-3)$.

\subsection{${\cal W}_3$ null-vector equations}

Let us first briefly explain why we try to relate conformal field theories with an $\widehat{s\ell_3}$ symmetry at level $k$ to theories with a ${\cal W}_3$ symmetry at central charge $c=2+24(b+b^{-1})^2$ where
\bea
\boxed{ b^2=\frac{1}{k-3}\ . }
\eea
A theory with an $\widehat{s\ell_3}$ symmetry like the $s\ell_3(\R)$ WZW model can be written in terms of eight quantum fields, as $s\ell_3$ is eight-dimensional. However, affine $\widehat{s\ell_3}$ highest-weight representations are parametrized by just two numbers, namely the two components of the $s\ell_3$ spin $j$. This suggests that the non-trivial dynamics of the theory really take place in a two-dimensional space, where $j$ would play the role of the momentum. There exists such an $s\ell_3$-based theory which involves just two interacting quantum fields: the conformal $s\ell_3$ Toda theory, which has a ${\cal W}_3$ symmetry algebra. The correct parameter $b$ for this algebra is suggested by the Drinfeld-Sokolov reduction, which realizes ${\cal W}_3$ as a kind of coset of the $\widehat{s\ell_3}$ algebra. 

\paragraph{${\cal W}_3$ algebra and primary fields.}

Referring to the review article \cite{bs92} for more details, we recall that the ${\cal W}_3$ algebra is spanned by the modes of the fields $T(z)=\sum_{n\in\Z} L_n z^{-n-2}$ and $W(z)=\sum_{n\in\Z}W_n z^{-n-3}$. Let us write the defining relations of the ${\cal W}_3$ algebra in the form of commutation relations for the modes $L_n,W_n$ rather than operator product expansions for the fields $T(z),W(z)$, as this form is more convenient for finding null vectors in representations:
\bea
[L_m,L_n]&=&(m-n)L_{m+n}+\frac{c}{12}m(m^2-1)\delta_{m+n,0} \ ,
\\ {}
[L_m,W_n]&=&(2m-n)W_{m+n}\ ,
\\ {}
[W_m,W_n]&=&\frac{(22+5c)}{48}\frac{c}{360} m(m^2-1)(m^2-4)\delta_{m+n,0} 
\nn \\
& & + \frac{(22+5c)}{48}\frac{m-n}{15}(m^2+n^2-\tfrac12 mn-4)  L_{m+n} 
+\frac13(m-n)\Lambda_{m+n}\ ,
\label{walg}
\eea
where we introduce, using the normal ordering $:L_mL_n:=L_mL_n$ if $m\leq n$,
\bea
\Lambda_m =\sum_{n\in \Z}  : L_nL_{m-n} :+\frac15 x_m L_m \ \ \ \ \ {\rm with}\ \ \ \left\{\begin{array}{l}  x_{2\ell}=(1+\ell)(1-\ell)\\ x_{2\ell+1} =(\ell+2)(1-\ell)\end{array}\right.\ .
\eea
A primary fields $V_\al$ of the ${\cal W}_3$ algebra of momentum $\al$, conformal dimension $\Delta_\al$ and charge $q_\al$ is defined by its operator product expansions with $T(z)$ eq. (\ref{tva}) and $W(z)$:
\bea
W(z)V_\al(w) = \frac{q_\al V_\al(w)}{(z-w)^3} +\frac{W_{-1}V_\al(w)}{(z-w)^2} +\frac{W_{-2}V_\al(w)}{z-w} + {\cal O}(1)\ .
\eea
The momenta $\al$ now belong to the two-dimensional root space of the Lie algebra $s\ell_3$. A basis of this space is provided by the simple roots $e_1,\ e_2$ whose scalar products appear in the Cartan matrix $\bsm (e_1,e_1) & (e_1,e_2) \\ (e_2,e_1) & (e_2,e_2) \esm = \bsm 2 & -1 \\ -1 & 2 \esm$. We may also use the dual basis $\om_1=\frac23 e_1+\frac13 e_2,\ \om_2=\frac13 e_1 + \frac23 e_2$ such that $(e_i,\om_j) =\delta_{ij}$. We decompose the momenta along this dual basis: $\al = \al_1 \om_1+\al_2 \om_2$, and we introduce the vector $Q=(b+b^{-1})(e_1+e_2)$. The conformal dimension and charge are parametrized in terms of the momentum as
\bea
\Delta_\al &=&\frac12(\al,2Q-\al)\ ,
\\
q_\al &=& \frac{i}{27} [\al_1-\al_2][2\al_1+\al_2-3(b+b^{-1})] [\al_1+2\al_2-3(b+b^{-1})]\ .
\eea

\paragraph{${\cal W}_3 $ degenerate fields.} 

Let us now justify the choice of the field $V_{-b^{-1}\om_1}$ in the correlator $\tilde{\Omega}_n$ (\ref{tomn}) which appears in our conjecture. We wish $\tilde{\Omega}_n$ to obey third-order differential equations, which would correspond to the $s\ell_3$ KZ equations in Sklyanin variables. This suggests that we use 
the simplest non-trivial degenerate fields, which have null vectors at levels 1, 2 and 3. 
But there are actually four such degenerate fields, with $\al\in \{-b\om_1,-b\om_2,-b^{-1}\om_1,-b^{-1}\om_2\}$, whereas we want only one of them to appear in $\tilde{\Omega}_n$, because the 
original isospin variables are invariant under permutations of the Sklyanin variables.

By analogy with the $s\ell_2$ case, we focus on the fields $V_{-b^{-1}\om_1}$ and $V_{-b^{-1}\om_2}$, whose momenta go to zero in the critical level limit $k\rar 3$. They are related to the other two fields by the ${\cal W}_3$ algebra self-duality $b\rar b^{-1}$, which is however not an invariance of the $\widehat{s\ell_3}$ algebra. 
And they are related to each other by the Dynkin diagram automorphism $\om_1\lrar \om_2$ of $s\ell_3$, which acts on general primary fields $V_\al$ as $(\Delta_\al,q_\al)\rar (\Delta_\al,-q_\al)$. This symmetry does have a counterpart in the separation of variables for the $s\ell_3$ Gaudin model. The construction of the separated variables was indeed based on the introduction of an $s\ell_3$ Lax matrix $I(u)$ (\ref{iu}), so that $s\ell_3$ generators act in the fundamental representation. But we could alternatively have used the antifundamental representation, which is related to the fundamental by the Dynkin diagram automorphism. 
With our conventions,
our choice of the fundamental representation will turn out to correspond to the choice of the degenerate field $V_{-b^{-1}\om_1}$ of the ${\cal W}_3$ algebra. The three corresponding null-vector equations are \cite{wat94}
\bea
% with \alpha = -ib
& &\left[ i W_{-1} + \left(\tfrac{b}{2} +\tfrac{5}{6b}\right) L_{-1}\right] V_{-b^{-1}\om_1} = 0 \ ,
\label{nvu}
\\
& &\left[ i W_{-2} - \tfrac{2}{3b} L_{-2} - bL_{-1}^2 \right] V_{-b^{-1}\om_1} = 0 \ ,
\label{nvd}
\\
& &\left[ i W_{-3} -\left(\tfrac{b}{2}+\tfrac{1}{6b}\right) L_{-3} + b L_{-1}L_{-2} + b^3 L_{-1}^3 \right] V_{-b^{-1}\om_1} = 0\ .
\label{nvt}
\eea
The last null-vector equation implies that any correlation function with one degenerate field obeys $E_1\la   V_{-b^{-1}\om_1}(y)\prod_{i=1}^n V_{\al_i}(z_i) \ra=0$, where
\begin{multline}
% still \alpha = -ib
E_1\equiv  \ppt{y} + \frac{1}{b^2} \pp{y}\cdot \sum_{i=1}^n \left(\frac{1}{y-z_i}\pp{z_i} +\frac{\Delta_{\al_i}}{(y-z_i)^2}\right)
\\
+\left(\frac{1}{2b^2} +\frac{1}{6b^4}\right) \sum_{i=1}^n\left(\frac{1}{(y-z_i)^2}\pp{z_i} +\frac{2\Delta_{\al_i}}{(y-z_i)^3}\right) 
 \\ 
 +\frac{i}{b^3}\sum_{i=1}^n\left(\frac{W_{-2,(i)}}{y-z_i} 
+\frac{W_{-1,(i)}}{(y-z_i)^2} +\frac{q_{\al_i}}{(y-z_i)^3} \right) \ .
\end{multline}
This may be compared with eq. (\ref{sepkz}), which is formally similar, or even identical if the term with coefficient $\frac{1}{6b^4}$ is absorbed into the other terms by redefining $W_{-1,(i)}$ and $q_{\al_i}$. Like in the $s\ell_2$ case, the meaning of this formal similarity is not clear. 

Now the equations obeyed by correlation functions with several degenerate fields like $\tilde{\Omega}_n$ eq. (\ref{tomn}) are significantly more complicated than $E_1$, because eliminating $W_{-1},W_{-2}$ descendents of the degenerate fields requires the use of the first two null-vector equations (\ref{nvu},\ref{nvd}). Still denoting $\{y_a\}=\{y,y_b\}$, we obtain the equation $E_2\tilde{\Omega}_n=0$ with
\begin{multline}
 E_2\equiv E_1 + \frac{1}{b^2}\sum_b \frac{1}{y-y_b}\ppd{y_b} +\frac{1}{b^2} \pp{y} \cdot \sum_b\left(\frac{1}{y-y_b}\pp{y_b} +\frac{\Delta_{-b^{-1}\om_1}}{(y-y_b)^2}\right) 
 \\
 +\frac{2}{3b^4}\sum_{b,i} \frac{1}{(y-y_b)(y_b-z_i)}\left(\pp{z_i} +\frac{\Delta_{\al_i}}{y_b-z_i}\right) +\frac{2}{3b^4}\sum_{b\neq c} \frac{1}{(y-y_b)(y_b-y_c)}\left(\pp{y_c} +\frac{\Delta_{-b^{-1}\om_1}}{y_b-y_c}\right)
 \\
 -\frac{2}{3b^4} \sum_b\frac{1}{(y-y_b)^2}\left(\pp{y_b}+\pp{y}\right) +\left(\left(\frac{1}{b^2}+\frac{1}{b^4}\right)\Delta_{-b^{-1}\om_1} +\frac{i}{b^3} q_{-b^{-1}\om_1}\right)\sum_b\frac{1}{(y-y_b)^3}\ ,
\label{eteo}
\end{multline}
where
\bea
\Delta_{-b^{-1}\om_1} = -1-\frac{4}{3b^2} \scs q_{-b^{-1}\om_1} = -\frac{i}{27b^3}(4+3b^2)(5+3b^2)\ .
\eea

\paragraph{Relating ${\cal W}_3$ momenta to $\widehat{s\ell_3}$ spins.}

In order to compare the equation $E_2\tilde{\Omega}_n=0$ with the KZ equations in Sklyanin variables (\ref{sepmod}), we should specify how we relate $\widehat{s\ell_3}$ primary fields $\Phi^j(\mu|z)$ to ${\cal W}_3$ primary fields $V_\al(z)$. We are looking for a relation between $\al$ and $j$ which translates into a simple relation between $(\Delta_\al,q_\al)$ and $(\Delta^J_j,q_j^J)$. We propose
\bea
\boxed{\al = -bj +b^{-1}(e_1+e_2)\ \Rightarrow\ \left\{\begin{array}{l} \Delta_\al=\Delta^J_j+2+b^{-2} \\ q_\al=q^J_j \end{array}\right.\ , }
\label{alj}
\eea
where we use the following expressions for $(\Delta^J_j,q^J_j)$ defined in eqs. (\ref{ljz}) and (\ref{wzj}) 
\bea
\Delta_j^J&=& -\frac{1}{k-3}\frac12 (j,j+2e_1+2e_2)\ ,
\\
q_j^J &=& \frac{1}{(k-3)^\frac32} \frac{i}{27}   [j_1-j_2]\left[2(j_1+1)+(j_2+1)\right]\left[(j_1+1)+2(j_2+1)\right]\ ,
\eea
where the components $(j_1,j_2)$ of the spin $j$ are defined as $j=j_1\om_1+j_2\om_2$. 
Notice that our relation between $\al$ and $j$ maps the principal unitary series of $s\ell_3$ representations $j\in -e_1-e_2+i\R^2$ to the ${\cal W}_3$ representations which appear in the physical spectrum of conformal $s\ell_3$ Toda theory \cite{fl07c} $\al \in Q+i\R^2$. Such choices of $\al$ or $j$ lead to real values of $(\Delta,q)$ if $k>3$.

% Relation with hamiltonian reduction?

However, there does not need to be any relation between the $\widehat{s\ell_3}$ creation operators $W_{-1}^J,W_{-2}^J$ and their ${\cal W}_3$ counterparts $W_{-1},W_{-2}$. While relating $L_{-1}^J=\ppl{z} $ to $L_{-1}=\pp{z} $, though difficult in practice, is in principle a simple matter of performing the change of variables, there is apparently no principle which would determine how $W_{-1}^J,W_{-2}^J$ would behave through the change of variables. 
This is why we work modulo ${\cal F}_2(y)$, ignoring the non-differential terms which involve such operators, and being left with differential equations.
Now the presence of degenerate fields in correlation functions of ${\cal W}_3$ fields does not necessarily lead to differential equations, a fact which makes conformal $s\ell_3$ Toda theory much more complicated than Liouville theory \cite{fl07c}. Differential equations actually appear provided the number of degenerate fields is large enough.
We are inserting $3n-6$ degenerate fields $V_{-b^{-1}\om_1}$ together with the $n$ generic fields $V_{\al_i}$, which is enough for eliminating the $2n$ terms $W_{-1,(i)},W_{-2,(i)}$ and being left with $n-6$ differential equations.

\paragraph{Twisting ${\cal W}_3$ null-vector equations.}

Finally, we should determine the twist factor $\Theta_n$ which appears in the conjecture (\ref{omom}), so as to be able to compute
\bea
E_3\equiv \Theta_n E_2 \Theta_n^{-1}\ \ \ {\rm such\ that} \ \ \ E_3\cdot \Theta_n\tilde{\Omega}_n=0\ .
\eea
The values of the parameters $\lambda,\nu$ can be derived as in the $s\ell_2$ case. Requiring continuity of $\Theta_n \tilde{\Omega}_n$ at $y_a=y_b$ implies $\lambda=2\Delta_{-b^{-1}\om_1}-\Delta_{-2b^{-1}\om_1} = \frac{2}{3b^2}$, and requiring that the conjecture (\ref{omom}) holds in the case $n=2$ implies $\nu=2\Delta_\al-2\Delta^J_j=\frac{2}{b^2}+4$, see eq. (\ref{alj}). Notice however that this only determines $\nu$ up to $b$-independent terms, as the unknown $b$-independent kernel $K$ may also contribute.

These constraints leave the parameter $\mu$ arbitrary. We will obtain an ansatz for $\mu$, and confirm the values of $\lambda$ and $\nu$, by generalizing the relation (\ref{thav}) between  $\Theta_n$ and free field correlation functions which was observed in the $s\ell_2$ case. In the $s\ell_3$ case the analogous relation is 
\bea
|\Theta_n|^{-2} = \la \prod_{a=1}^{3n-6} V_{-b^{-1}\om_1}(y_a) \prod_{i=1}^n V_{b^{-1}(e_1+e_2)}(z_i)\ra^{free}\ .
\label{thaw}
\eea
This ansatz leads to the values
\bea
\boxed{\lambda=\frac{2}{3b^2} \scs \mu=-\frac{1}{b^2} \scs \nu = \frac{2}{b^2}\ .}
\label{lmnw}
\eea
These values will turn out to be the only ones such that, modulo ${\cal F}_2(y)$, the only non-differential terms in $E_3$ are of the type $\frac{c_i}{(y-z_i)^3}$. This is a rather non-trivial requirement as many non-differential terms can potentially appear (cf Appendix \ref{apid}).
Working modulo ${\cal F}_2(y)$ eq. (\ref{fty}), and using the relation (\ref{alj}) between $\widehat{s\ell_3}$ and ${\cal W}_3$ representation data, we indeed compute
\bea
\boxed{
E_3  \sim \ppt{y} +  \frac{1}{b^2} D_2 +\frac{1}{b^4} D_1  + \frac{1}{b^2}\sum_{i=1}^n \frac{\Delta_{j_i}^J}{(y-z_i)^2}\pp{y} +\sum_{i=1}^n \frac{ \frac{i}{b^3} q^J_{j_i} -\frac{1}{b^2} \Delta^J_{j_i}}{(y-z_i)^3} \ ,}
\label{et}
\eea
where we introduced two differential operators $D_1$ and $D_2$ of respective orders $1$ and $2$, which depend neither on the field momenta $\al_i$ nor on the model parameter $b$, 
\bea
D_1&\equiv & -\sum_i\frac{1}{(y-z_i)^2}\pp{y} +2\left(\sum_i\frac{1}{y-z_i}\right)^2 \pp{y}
\nn
\\
 & & +3\sum_i\frac{1}{y-z_i}\sum_b\frac{1}{y-y_b}\left(\pp{y_b} -\pp{y} \right) 
 -2\sum_{b\neq c}\frac{1}{y-y_b}\frac{1}{y_b-y_c}\left(\pp{y_b} - \pp{y}\right)
 \ ,
\label{diop}
\\
D_2&\equiv & \sum_i \frac{1}{y-z_i} \pp{y} \left(\pp{z_i} +3 \pp{y} \right)+ \sum_b \frac{1}{y-y_b} \left(\pp{y_b} -\pp{y} \right) \left(\pp{y_b} +2\pp{y} \right) 
\nn
\\ & & 
+\sum_b\frac{1}{(y-y_b)^2} \pp{y}
 \ .
\eea

\subsection{Comparing $s\ell_3$ Knizhnik--Zamolodchikov equations with ${\cal W}_3$ null-vector equations\label{seccom}}

We are now in a position to test the conjecture (\ref{omom}) by comparing the KZ equations in Sklyanin variables (\ref{sepmod}), which apply to ${\cal K}^{-1}\Omega_n$, with the twisted ${\cal W}_3$ null-vector equations (\ref{et}), which apply to $\Theta_n \tilde{\Omega}_n$. We will first do the comparison for general values of $b$, and then explain in more detail what happens in the particular limits $b\rar \infty$ and $b\rar 0$. 

\paragraph{The comparison for general $b$.} 

To start with, the non-differential terms agree. This is actually a very non-trivial statement, as we started with complicated non-differential terms in eq. (\ref{eteo}) an then generated more terms by twisting with $\Theta_n$. The freedoms to choose the three parameters $\lambda,\mu,\nu$ of $\Theta_n$ and to ignore terms belonging to ${\cal F}_2(y)$ is a priori not sufficient to ensure the dozens of required cancellations, which nevertheless occur as can be seen in explicit calculations. These calculations use some helpful identities which are gathered in Appendix \ref{apid}. The existence of a simple twist which simplifies the differential equations obeyed by correlation functions involving many identical degenerate fields might well be a general phenomenon in conformal field theory, as we now see that it happens for the simplest degenerate field in theories  with ${\cal W}_3$ symmetry, in addition to the already known cases of the two simplest degenerate fields in theories with Virasoro symmetry \cite{sto00,rib08}.

Let us then examine the term $\frac{1}{b^2} D_2$ in eq. (\ref{et}). Agreement with the corresponding term in eq. (\ref{sepmod}) would occur provided 
\bea
\pp{y}\cdot \sum_i\frac{1}{y-z_i}{\cal K}^{-1}\ppl{z_i}{\cal K} \overset{?}{\sim} D_2\ .
\label{tof}
\eea
It seems technically challenging to check this identity. But remember that our inability to explicitly perform Sklyanin's change of variables for $\ppl{z_i}$ does not contaminate the other terms in our equations, as we do know that the change of variables must be independent from the parameter $b=(k-3)^{-\frac12}$. 

Let us now examine the term $\frac{1}{b^4} D_1$. We would like this term to vanish modulo ${\cal F}_2(y)$, as no such term is present in eq. (\ref{sepmod}). However, it is rather obvious that $D_1$ does not belong to ${\cal F}_2(y)$, although it has quite a few remarkable properties. This is explained in detail in the Appendix \ref{appdo}.
As a result, the conjecture cannot hold for general values of $b$.

\paragraph{The critical level limit $b\rar \infty$.}

We notice that the term $\frac{1}{b^4} D_1$, which is responsible for the failure of our conjecture, vanishes in the $b\rar \infty$ limit. Therefore, the conjecture has better chances to hold in that limit. To completely prove that it does, we still need to clear one subtlety with the term $\frac{1}{b^2} D_2$. This term seems to vanish in the $b\rar \infty$ limit but actually it does not. This is because near $b\rar \infty$ our correlation functions do not have finit limits. Rather, the Toda correlation function $\tilde{\Omega}_n = \la \prod_a V_{-b^{-1}\om_1}(y_a) \prod_i V_{-bj_i+b^{-1}(e_1+e_2)}(z_i)\ra$ involves ``heavy'' fields $V_{-bj_i+b^{-1}(e_1+e_2)}(z_i)$ whose momenta grow as $b$. On general grounds (see for instance \cite{fl07c}), it is therefore expected that $\tilde{\Omega}_n\underset{b\rar \infty}{\sim} e^{b^2 S(\{z_i\})} T_n$ where $S$ and $T_n$ are $b$-independent functions, and $S$ depends only on $\{z_i\}$ and not on $\{y_a\}$. The differential operator $\frac{1}{b^2} D_2$, which contains derivatives with respect to $z_i$, may yield a finite contribution when such derivatives act on $e^{b^2 S(\{z_i\})}$.

We should therefore check whether eq. (\ref{tof}) holds to the leading order in $b^2$ when acting on functions of the type $e^{b^2 S(\{z_i\})} T_n$. This is actually the case, because the only term in $D_2$ with $z_i$-derivatives is $\pp{y}\cdot\sum_i\frac{1}{y-z_i}\pp{z_i}$, and ${\cal K}^{-1}\ppl{z_i}{\cal K}\ S(\{z_i\}) =\pp{z_i} S(\{z_i\})$.
This completes the proof of the conjecture (\ref{omom}) in the critical level limit $b\rar \infty \Leftrightarrow k\rar 3$.

Notice that this $b\rar \infty$ limit is not sensitive to the twist function $\Theta_n$. This is because the exponents $\lambda,\mu,\nu$ (\ref{lmnw}) vanish in this limit so that $\Theta_n \underset{b\rar \infty}{\rar} 1$.

\paragraph{The minisuperspace limit $b\rar 0$.}

In this limit, the discrepant term $\frac{1}{b^4} D_1$, which is responsible for the failure of the conjecture (\ref{omom}) for general $b$, grows larger. We may therefore obtain some insights on the reasons for this failure.  

As in the $s\ell_2$ case, we will consider full correlation functions (with both holomorphic and antiholomorphic dependences) and use path-integral reasonings in $s\ell_3$ Toda theory. For full correlation functions, the conjecture reads $\Omega_n \sim {\cal K}{\cal \bar{K}} |\Theta_n|^2\tilde{\Omega}_n$. As in the $s\ell_2$ case, the transformation ${\cal K}$ is $b$-independent, $\Omega_n$ is expected to have a finite limit, and the Toda correlation function $\tilde{\Omega}_n$ behaves as $\tilde{\Omega}_n \underset{b\rar 0}{\sim} R_n \la \prod_{a=1}^{3n-6} V_{-b^{-1}\om_1}(y_a) \prod_{i=1}^n V_{b^{-1}(e_1+e_2)}(z_i)\ra $ where $R_n$ is $b$-independent. 

Therefore $\tilde{\Omega}_n$ simplifies in the $b\rar 0$ limit but, in contrast to the $s\ell_2$ case, its leading behaviour does not reduce to a free field correlation function. This is because the simplified correlation function $\la \prod_{a=1}^{3n-6} V_{-b^{-1}\om_1}(y_a) \prod_{i=1}^n V_{b^{-1}(e_1+e_2)}(z_i)\ra$ does not obey momentum conservation, given the value $2Q=2(b+b^{-1})(e_1+e_2)$ of the background charge in $s\ell_3$ Toda theory. However, momentum conservation can be restored by inserting $n-2$ screening operators $V_{b^{-1}e_1}$. (See \cite{fl07c} for similar reasonings and calculations in $s\ell_3$ Toda theory.) Thus, 
\bea
\tilde{\Omega}_n &\underset{b\rar 0}{\sim}&  R_n\la \prod_{a=1}^{3n-6} V_{-b^{-1}\om_1}(y_a) \prod_{i=1}^n V_{b^{-1}(e_1+e_2)}(z_i) \prod_{\ell=1}^{n-2} \int d^2x_\ell\ V_{b^{-1}e_1}(x_\ell) \ra^{free}\ .
\eea
This free correlation function is the product of the free correlation function (\ref{thaw}), which we took as our ansatz for $|\Theta_n|^2$, and an integral over $x_\ell$, leading to 
\bea
|\Theta_n|^2\tilde{\Omega}_n &\underset{b\rar 0}{\sim}& R_n \int \prod_\ell d^2x_\ell\ \prod_{a,\ell} |x_\ell-y_a|^{\frac{2}{b^2}} \prod_{i,\ell} |x_\ell-z_i|^{-\frac{2}{b^2}} \prod_{\ell\neq\ell'} |x_\ell-x_{\ell'}|^{-\frac{4}{b^2}}\ .
\eea
The integral in this formula is expected to be dominated by a saddle point, where the $x_\ell$s are solutions of
\bea
2\sum_{\ell'\neq \ell}\frac{1}{x_\ell-x_{\ell'}} +\sum_i\frac{1}{x_\ell-z_i} -\sum_a\frac{1}{x_\ell-y_a} = 0 \ .
\eea
(Curiously, these are the Bethe equations for the $s\ell_2$ Jaynes-Cummings-Gaudin model at infinite coupling and with spins $\pm \frac12$ \cite{bt07}.) The dominant behaviour of the integral is expected to be of the form
$|\theta_n|^{\frac{2}{b^2}}$ as $b\rar 0$, with $\theta_n$ a $b$-independent quantity. This $|\theta_n|^{\frac{2}{b^2}}$ factor contradicts the existence of a finite limit for $|\Theta_n|^2\tilde{\Omega}_n$ as $b\rar 0$, which follows from the conjecture. 

One may be tempted to modify the conjecture by adding a factor $\theta_n^{-\frac{1}{b^2}}$ to the twist function $\Theta_n$. This would not only correct the leading behaviour in the $b\rar 0$ limit, but also make the conjecture compatible with global conformal symmetry. We have not mentioned global conformal symmetry until now because this subject is independent from the differential equations in terms of which the conjecture was formulated. It is however easy to see that the conjecture is incompatible with the behaviour of correlation functions under scaling transformations $(z_i,y_a)\rar (\lambda z_i,\lambda y_a)$, except in the $b\rar \infty$ limit.

However, adding the factor $\theta_n^{-\frac{1}{b^2}}$ would spoil the agreement between most terms of the KZ equations (\ref{sepmod}) and the twisted ${\cal W}_3$ null-vector equations (\ref{et}), in particular the terms depending on the spins $j_i$. The modified conjecture would only hold at the level of the $j_i$-independent dominant factors in the $b\rar 0$ limit, which would not be interesting.

\section{Conclusion}

The comparison of $s\ell_3$ KZ equations in Sklyanin variables (\ref{sepmod}) with ${\cal W}_3$ null-vector equations (\ref{et}) does not support the conjecture (\ref{omom}) in its general form. Nevertheless, the KZ equations are very similar to the null-vector equations: many terms agree nontrivially, and the disagreement is confined to a term which does not depend on the spins $j_i$ of the fields. This remarkable quasi-agreement makes it unlikely that a full agreement can be obtained by modifying the conjecture.

In the critical level limit $k\rar 3\Leftrightarrow b\rar \infty$, the disagreement disappears and the conjecture (\ref{omom}) is true. This limit plays an important role in the Langlands correspondence \cite{fre05}, which might possibly explain why the conjecture (\ref{omom}) holds for $s\ell_2$ and not for $s\ell_3$, and why in the $s\ell_3$ case it holds only in the critical level limit. Another hopeful source of insights is the recent work on conformal Toda theories \cite{fl07c}, where the $s\ell_{N\geq 3}$ cases are understood to be qualitatively different from the $s\ell_2$ case. Of course, we already pointed out a significant qualitative difference, namely the failure of the $\widehat{s\ell_3}$ cubic field $W^J(z)$ (\ref{wz}) to obey the ${\cal W}_3$ algebra. It is not clear how this is related to our problem. 

Our results in the $s\ell_3$ case lead to natural conjectures in $s\ell_{N>3}$ cases, where we expect the KZ equations in Sklyanin variables to agree with ${\cal W}_N$ null-vector equations only in the critical level limit $k\rar N$.
Let us tentatively perform a counting of equations. There are $\frac12 N(N-1)$ isospin variables on the lhs of eq. (\ref{omom}), and on the rhs we expect $\frac12 N(N-1)(n-2)$ Sklyanin variables $y_a$ plus $N(N-1)$ extra variables, which may be collectively included in the symbol $\ev$.
Differential equations for the $s\ell_N$ Toda correlation function which generalizes $\tilde{\Omega}_n$ are obtained by eliminating $\frac12 (N-2)(N+1)n$ non-differential terms from the $\frac12 N(N-1)(n-2)$ null-vector equations. Thus, we have $n-N(N-1)$ differential equations. When it comes to ${\cal K}\Theta_n\tilde{\Omega}_n$, we should presumably add an equation for each one of the extra variables, reaching $n$ differential equations. This precisely the number of KZ equations for the lhs $\Omega_n$ of eq. (\ref{omom}). In addition, we have the same number of global Ward identities on both sides of eq. (\ref{omom}), namely $N^2-1$.

\vfill

\pagebreak

\appendix

\zeq\section{A few technical results}

\subsection{Helpful identities \label{apid}}

The following identities are used in computing the non-differential terms of the operator $E_3\equiv \Theta_n E_2\Theta_n^{-1}$ eq. (\ref{et}). Some identites are written modulo terms in ${\cal F}_2(y)$ (\ref{fty}), as indicated by the relation sign $\sim$. All identities are proved by elementary manipulations, using observations of the type $\frac{1}{(y-z_i)^2} \sum_b\frac{1}{y_b-z_i} = \frac{1}{(y-z_i)^2}\left(\sum_a\frac{1}{y_a-z_i}-\frac{1}{y-z_i}\right) \sim -\frac{1}{(y-z_i)^3}$.
\bea
\left(\sum_i \frac{1}{y-z_i}\right)^3 &\sim & \sum_i\frac{1}{y-z_i}\sum_j\frac{1}{(y-z_j)^2} \sim \sum_i\frac{1}{(y-z_i)^3}\ ,
\\
\sum_b \frac{1}{y-y_b} \frac{1}{(y_b-z_i)^2} &\sim & -\frac{2}{(y-z_i)^3}+\frac{1}{(y-z_i)^2}\sum_b\frac{1}{y-y_b}\ ,
\\
\sum_b\frac{1}{y-y_b} \left(\sum_i\frac{1}{y_b-z_i}\right)^2 &\sim & \sum_i\frac{-2}{(y-z_i)^3} + \sum_b\frac{1}{y-y_b}\left(\sum_i \frac{1}{y-z_i}\right)^2 \ ,
\\
\sum_{bij} \frac{1}{y-y_b}\frac{1}{y-z_j}\frac{1}{y_b-z_i} &\sim & \sum_i\frac{-1}{(y-z_i)^3} +\sum_b\frac{1}{y-y_b} \left(\sum_i\frac{1}{y-z_i}\right)^2 \ ,
\\
\sum_b \frac{1}{y-y_b}\frac{1}{y_b-z_i}\frac{1}{y_b-z_j} &\sim& \frac{1}{(y-z_i)(y-z_j)}\sum_b \frac{1}{y-y_b} 
\eea
\bea
\sum_b\frac{1}{(y-y_b)^2} \frac{1}{y_b-z_i} &\sim & -\frac{1}{(y-z_i)^3} +\frac{1}{(y-z_i)^2}\sum_b\frac{1}{y-y_b} +\frac{1}{y-z_i} \sum_b\frac{1}{(y-y_b)^2} \ ,
\\
\sum_{b\neq c}\frac{1}{y-y_b}\frac{1}{y_b-y_c}\frac{1}{y_b-z_i} &\sim & -\frac{1}{(y-z_i)^3}+\frac12\frac{1}{y-z_i}\left[\left(\sum_b\frac{1}{y-y_b}\right)^2-
\sum_b\frac{1}{(y-y_b)^2} \right]\ ,
\\
\sum_{b\neq c}\frac{1}{y-y_b}\frac{1}{y_b-y_c}\frac{1}{y_c-z_i} &\sim & 
\frac{2}{(y-z_i)^3}+\frac12\frac{1}{y-z_i}\left[\left(\sum_b\frac{1}{y-y_b}\right)^2-
\sum_b\frac{1}{(y-y_b)^2} \right]
\\
& & -\frac{1}{(y-z_i)^2} \sum_b\frac{1}{y-y_b} +\frac{1}{y-z_i} \sum_b\frac{1}{y-y_b}\sum_c\frac{1}{y_c-z_i} \ ,
\eea
\bea
\sum_{b\neq c}\frac{1}{(y-y_b)^2}\frac{1}{y_b-y_c} &=&\sum_b\frac{1}{(y-y_b)^2}\sum_c\frac{1}{y-y_c}-\sum_b\frac{1}{(y-y_b)^3}\ ,
\\
\sum_{\begin{smallmatrix} b\neq c \\ d\neq c \end{smallmatrix}} \frac{1}{y-y_b}\frac{1}{y_b-y_c}\frac{1}{y_c-y_d} &=& -\sum_{b\neq c}\frac{1}{y-y_b}\frac{1}{(y_b-y_c)^2} +\frac16 \sum_{b\neq c\neq d} \frac{1}{y-y_b}\frac{1}{y-y_c}\frac{1}{y-y_d}\ ,
\\
\sum_{\begin{smallmatrix} c\neq b \\ d\neq b \end{smallmatrix}} \frac{1}{y-y_b}\frac{1}{y_b-y_c}\frac{1}{y_b-y_d} &=& \sum_{b\neq c}\frac{1}{y-y_b}\frac{1}{(y_b-y_c)^2} +\frac13 \sum_{b\neq c\neq d} \frac{1}{y-y_b}\frac{1}{y-y_c}\frac{1}{y-y_d}\ ,
\\
\sum_{b\neq c\neq d} \frac{1}{y-y_b}\frac{1}{y-y_c}\frac{1}{y-y_d} &=& \left(\sum_b\frac{1}{y-y_b}\right)^3+\sum_b\frac{2}{(y-y_b)^3}- 3\sum_b\frac{1}{y-y_b}\sum_c\frac{1}{(y-y_c)^2}\ .
\eea

\subsection{A characterization of ${\cal F}_2(y)$ \label{apf}}

Here we will justify the characterisation (\ref{chft}) of the space ${\cal F}_2(y)$ defined in eq. (\ref{fty}). 

For pedagogical reasons we will begin with the simpler problem of characterizing the space of permutation-symmetric functions of $m$ variables $\{y_a\}$. More precisely, given a function $f(t,\{y_a\})$ which is permutation-symmetric in $\{y_a\}$, depends on an additional variable $t$, and is regular at $t=y_a$, we want to determine whether $f(y,\{y_a\})$ is actually permutation-symmetric although it apparently depends on $y$. 
This amounts to determining whether $f(y_{a'},\{y_a\})$ actually depends on the choice of $a'$. If it does not, then for any polynomial $P(t)$ of degree $m-2$ we have 
\bea
\sum_{a'} \oint_{y_{a'}}dt\ \frac{P(t) f(t)}{\prod_a(t-y_a)} = f(y) \sum_{a'}\oint_{y_{a'}}dt\ \frac{P(t)}{\prod_a(t-y_a)} = f(y)\oint_\infty dt\ \frac{P(t)}{\prod_a(t-y_a)} = 0\ .
\eea
So we have transformed the $m-1$ conditions $f(y_1)=f(y_2)=\cdots = f(y_m)$ into the condition $\sum_{a'} \oint_{y_{a'}}dt\ \frac{P(t) f(t)}{\prod_a(t-y_a)}=0$, which can then be evaluated by moving the integration contours, if the analytic properties of $f(t)$ permit.

Let us apply a similar reasoning to the characterization of ${\cal F}_2(y)$. If $f(y)\in {\cal F}_2(y)$, for instance $f(y)=\frac{1}{(y-z_{i_0})^2}\tilde{f}(y)$ where $\tilde{f}(y)$ is actually permutation-symmetric, then given any polynomial $P(t)$ of degree $n-7$ we have
\bea
\sum_{a'} \oint_{y_{a'}}dt\ P(t) \frac{\prod_{i=1}^n (t-z_i)^2}{\prod_{a=1}^{3n-6}(t-y_a)} f(t)  = \tilde{f}(y) \oint_\infty dt\ P(t) \frac{\prod_{i\neq i_0} (t-z_i)^2}{\prod_a(t-y_a)}=0\ .
\eea
Thus, to know whether $f(y)\in {\cal F}_2(y)$, we only need to evaluate the left hand-side of this equality. To do this we can use the assumed analytic properties of $f(t)$: namely, that it is meromorphic with singularities only at $t=z_i$, and goes to zero as $t\rar \infty$. This implies 
\bea
\sum_{a'} \oint_{y_{a'}}dt\ P(t) \frac{\prod_i (t-z_i)^2}{\prod_a(t-y_a)} f(t) = - \sum_{i=1}^n \oint_{z_i}dt\ P(t) \frac{\prod_i (t-z_i)^2}{\prod_a(t-y_a)} f(t)\ ,
\eea
which proves $f(y)\in {\cal F}_2(y)\Rightarrow \la P,f\ra =0$ as in eq. (\ref{chft}). The reverse implication follows from a simple counting of variables: the space of polynomials of degree $n-7$ has dimension $n-6$, which is precisely the number of constraints which we expect for characterizing the space ${\cal F}_2(y)$. 

\subsection{Study of the differential operator $D_1$\label{appdo}}

As explained in Section \ref{seccom}, our conjecture (\ref{omom}) implies the relation $D_1\overset{?}{\sim} 0$ or equivalently $D_1\overset{?}{\in} {\cal F}_2(y)$, where $D_1$ is the first-order differential operator written explicitly in eq. (\ref{diop}). Here we provide a rigorous argument that this relation is not true, which implies that the conjecture cannot hold for general values of the parameter $b$. 

To start with, let us reduce the study of the first-order differential operator $D_1$ to the study of mere functions. The operator $D_1$, like all our differential equations, is assumed to act on functions which are symmetric under permutations of the $3n-6$ variables $\{y_a\}$. The space of such functions
is algebraically generated by the $3n$ functions
\bea
\rho_i\equiv \sum_a \log(y_a-z_i) \scs \sigma_i\equiv \sum_a \frac{1}{y_a-z_i}\scs \tau_i\equiv \sum_a \frac{1}{(y_a-z_i)^2}\scs  i=1\cdots n.
\eea
Therefore, $D_1\sim 0 \Leftrightarrow D_1\rho_i\sim D_1\sigma_i\sim D_1\tau_i\sim 0$. 
Direct calculations show
\bea
D_1\rho_i &\sim & 0\ ,
\\
D_1 \sigma_i &\sim & -\frac{1}{(y-z_i)^4} +\frac{\sigma_i}{(y-z_i)^3} +\frac{2}{(y-z_i)^3}\sum_{j\neq i}\frac{1}{y-z_j}\ ,
\\
D_1\tau_i &\sim & -\frac{8}{(y-z_i)^5} +\frac{6\sigma_i}{(y-z_i)^4} +\frac{10}{(y-z_i)^4}\sum_{j\neq i} \frac{1}{y-z_j} +\frac{4\tau_i}{(y-z_i)^3} -\frac{2\sigma_i^2}{(y-z_i)^3}
\\ & & -\frac{6\sigma_i}{(y-z_i)^3}\sum_{j\neq i}\frac{1}{y-z_j} +\frac{2}{(y-z_i)^3}\sum_{j\neq i} \frac{1}{(y-z_j)^2} -\frac{4}{(y-z_i)^3}\left(\sum_{j\neq i} \frac{1}{y-z_j}\right)^2\ .
\eea
So $D_1\sigma_i$ and $D_1\tau_i$ do not manifestly vanish modulo ${\cal F}_2(y)$. Let us however study them further. They may be considered as values at $t=y$ of functions $f(t)=f(t,\{y_a\},\{z_i\})$ which are invariant under permutations of $\{y_a\}$ but depend on the additional variable $t$. 
Let us consider the space of such functions, which we in addition assume to be meromorphic in $t$ with no singularities besides $t=z_i$, and to go to zero as $t\rar \infty$. Let us moreover introduce the space ${\cal P}_{n-7}$ of polynomials $P(t)$ of degree $n-7$.
As we show in Appendix \ref{apf}, 
\bea
f(y)\in {\cal F}_2(y)\ \Leftrightarrow\ \forall P\in {\cal P}_{n-7}, \ \ \la P,f\ra \equiv\sum_{i=1}^n\oint_{z_i}dt\  P(t)\frac{\prod_{i=1}^n (t-z_i)^2}{\prod_{a=1}^{3n-6} (t-y_a)} f(t)=0\ .
\label{chft}
\eea
Then, explicit calculations yields
\bea
\la P, D_1\sigma_i\ra &=& 2\pi i\frac{\prod_{k\neq i} (z_i-z_k)^2}{\prod_a(z_i-y_a)} P'(z_i)\ ,
\\
\la P, D_1\tau_i\ra &=& 2\pi i\frac{\prod_{k\neq i} (z_i-z_k)^2}{\prod_a(z_i-y_a)}
\left[4P''(z_i) +\left(2\sigma_i+6\sum_{k\neq i}\frac{1}{z_i-z_k} \right)P'(z_i)\right]\ .
\eea
This explicitly demonstrates that $D_1\notin {\cal F}_2(y)$. 

However, $D_1$ still has remarkable properties with respect to the constant polynomial $P=1$, namely $\la 1,D_1\sigma_i\ra = \la 1,D_1\tau_i\ra =0$. These non-trivial identities sensitively depend on the general structure of $D_1$ and on the particular values of $\lambda,\mu,\nu$ which determine its coefficients. This implies that, whereas arbitrary differential operators belong to ${\cal F}_2(y)$ for $n\leq 6$, $D_1\in {\cal F}_2(y)$ for $n\leq 7$. The significance of these properties of $D_1$ is not clear. When combined with $D_1\rho_i\sim 0$, they suggest that $D_1\sim 0$ when applied to a special class of permutation-symmetric function of $y_a$ (and $z_i$), and one might wonder whether $\Theta_n \tilde{\Omega}_n$ actually belongs to this class. Given the freedom to choose $y\in \{y_a\}$, this would imply that $\tilde{\Omega}_n$ satisfies $n-6$ further differential equations. But $\tilde{\Omega}_n$ is not expected to satisfy any further differential equations besides the global Ward identities, whose number is $n$-independent. So the supposition $D_1\cdot \Theta_n \tilde{\Omega}_n\overset{?}{\sim} 0$ certainly fails for $n>7$, and so does our conjecture (\ref{omom}).

\acknowledgments{I wish to thank Alexander Chervov and Alexey Litvinov for interesting discussions and very helpful comments on this manuscript.
I am also grateful to Nicolas Cramp\'e, Vladimir Fateev, Philippe Roche, Volker Schomerus and Joerg Teschner for stimulating discussions.
I acknowledge the hospitality of DESY, Hamburg while part of this work was done.}

\bibliographystyle{JHEP-2}
%\bibliographystyle{morder}
%\bibliography{992}

\end{document}